\documentstyle[12pt]{article}

\def\hybrid{\topmargin 0pt      \oddsidemargin 0pt
        \headheight 0pt \headsep 0pt
        \voffset=-0.5cm
        \textwidth 6.25in       
        \textheight 9.5in       
        \marginparwidth 0.0in
        \parskip 5pt plus 1pt   \jot = 1.5ex}
\catcode`\@=11
\def\marginnote#1{}

\newcount\hour
\newcount\minute
\newtoks\amorpm
\hour=\time\divide\hour by60
\minute=\time{\multiply\hour by60 \global\advance\minute by-\hour}
\edef\standardtime{{\ifnum\hour<12 \global\amorpm={am}%
        \else\global\amorpm={pm}\advance\hour by-12 \fi
        \ifnum\hour=0 \hour=12 \fi
        \number\hour:\ifnum\minute<10 0\fi\number\minute\the\amorpm}}
\edef\militarytime{\number\hour:\ifnum\minute<10 0\fi\number\minute}

\def\draftlabel#1{{\@bsphack\if@filesw {\let\thepage\relax
   \xdef\@gtempa{\write\@auxout{\string
      \newlabel{#1}{{\@currentlabel}{\thepage}}}}}\@gtempa
   \if@nobreak \ifvmode\nobreak\fi\fi\fi\@esphack}
        \gdef\@eqnlabel{#1}}
\def\@eqnlabel{}
\def\@vacuum{}
\def\draftmarginnote#1{\marginpar{\raggedright\scriptsize\tt#1}}

\def\draftlabel#1{{\@bsphack\if@filesw {\let\thepage\relax
   \xdef\@gtempa{\write\@auxout{\string
      \newlabel{#1}{{\@currentlabel}{\thepage}}}}}\@gtempa
   \if@nobreak \ifvmode\nobreak\fi\fi\fi\@esphack}
        \gdef\@eqnlabel{#1}}
\def\@eqnlabel{}
\def\@vacuum{}
\def\draftmarginnote#1{\marginpar{\raggedright\scriptsize\tt#1}}

\def\draft{\oddsidemargin -.5truein
        \def\@oddfoot{\sl preliminary draft \hfil
        \rm\thepage\hfil\sl\today\quad\militarytime}
        \let\@evenfoot\@oddfoot \overfullrule 3pt
        \let\label=\draftlabel
        \let\marginnote=\draftmarginnote
   \def\@eqnnum{(\theequation)\rlap{\kern\marginparsep\tt\@eqnlabel}%
\global\let\@eqnlabel\@vacuum}  }

\def\numberbysection{\@addtoreset{equation}{section}
        \def\theequation{\thesection.\arabic{equation}}}

\def\underline#1{\relax\ifmmode\@@underline#1\else
        $\@@underline{\hbox{#1}}$\relax\fi}

\def\titlepage{\@restonecolfalse\if@twocolumn\@restonecoltrue\onecolumn
     \else \newpage \fi \thispagestyle{empty}\c@page\z@
        \def\thefootnote{\fnsymbol{footnote}} }

\def\endtitlepage{\if@restonecol\twocolumn \else  \fi
        \def\thefootnote{\arabic{footnote}}
        \setcounter{footnote}{0}}  
\relax

\numberbysection
\hybrid

\def\beq{\begin{equation}}
\def\eeq{\end{equation}}
\def\p{\partial}
\def\G{\Gamma}

\newtheorem{th}{Theorem}[section]

\newtheorem{cor}{Corollary}[section]
\newtheorem{lem}{Lemma}[section]

\def\square{\hfill
{\vrule height6pt width6pt depth1pt} \break \vspace{.01cm}}

\begin{document}

\begin{titlepage}
\title{
\rightline{\normalsize ITEP-TH-76/97}
Vacuum curves of elliptic $L$-operators and
representations of Sklyanin algebra}

\author{I.Krichever \thanks{Columbia University, 2990 Broadway,
New York, NY 10027, USA and
Landau Institute for Theoretical Physics,
Kosygina str. 2, 117940 Moscow, Russia; e-mail:
krichev@shire.math.columbia.edu}
\and A.Zabrodin
\thanks{Joint Institute of Chemical Physics, Kosygina str. 4, 117334,
Moscow, Russia and ITEP, 117259, Moscow, Russia; e-mail:
zabrodin@heron.itep.ru}}
\date{December 1997}

\maketitle

\begin{abstract}
An algebro-geometric approach
to representations of Sklyanin algebra is proposed.
To each 2$\times$2 quantum $L$-operator an algebraic
curve parametrizing its possible vacuum states is associated.
This curve is called the vacuum curve of the $L$-operator.
An explicit description of the vacuum curve for
quantum $L$-operators of the integrable spin chain
of $XYZ$ type with arbitrary spin $\ell$ is given.
The curve is highly reducible. For half-integer $\ell$
it splits into $\ell \!+\!\frac{1}{2}$ components
isomorphic to an elliptic curve.
For integer $\ell$ it splits into $\ell$
elliptic components and one rational component.
The action of elements of the $L$-operator to
functions on the vacuum curve leads to a new realization
of the Sklyanin algebra by difference operators
in two variables restricted to an invariant functional subspace.
\end{abstract}

\vfill

\end{titlepage}

\section{Introduction}

The Yang-Baxter equation
\beq
R^{23}(u-v) R^{13}(u) R^{12}(v) =
R^{12}(v) R^{13} (u) R^{23}(u-v)
\label{01}
\eeq
is a key relation of the
theory of quantum integrable
models\footnote{We use the following standard notation:
let $R$ be a linear operator
acting in the tensor product
${\bf C}^n\otimes {\bf C}^n$, then $R^{ij},\ i,j=1,2,3$
is an operator in the tensor product
${\bf C}^n\otimes {\bf C}^n\otimes C^n$
which acts as $R$ in the tensor product of $i$-th and $j$-th factors
of the triple tensor product and as
identity operator on the factor left.}.
Each solution of eq.\,(\ref{01}) generates
a hierarchy of integrable models.
The commutation relations for elements of
quantum $L$-operators of this hierarchy are given by
the "intertwining" equation
\beq
R^{23}(u-v) L^{13}(u) L^{12}(v) =
L^{12}(v) L^{13} (u) R^{23}(u-v)
\label{02}
\eeq
Here $L$ is an operator in the tensor product
${\bf C}^N\otimes {\bf C}^n$ and all the
factors in (\ref{02}) are operators in the tensor product
${\bf C}^N\otimes {\bf C}^n\otimes {\bf C}^n$.

Let $n=2$ in (\ref{01}); then the most general $R$-matrix
with elliptic dependence of the spectral parameter $u$
corresponds to the famous
8-vertex model (or, equivalently, to the $XYZ$ magnet):
\beq
R(u)=\sum_{a=0}^3 W_{a}(u+\eta )
\sigma_{a}\otimes \sigma_{a}
\label{03}
\eeq
Here $\sigma_a$ are Pauli matrices ($\sigma_0$ is
the unit matrix), $W_a (u)$
are functions of
$u$ with parameters $\eta$ and $\tau$:
\beq
W_a(u)={\theta_{a+1}(u|\tau )\over
\theta_{a+1}(\eta |\tau )}
\label{2}
\eeq
(the Jacobi $\theta$-functions are listed in the Appendix).

In \cite{Skl1},\,\cite{Skl2} Sklyanin reformulated the problem
of solving equation (\ref{02})
in terms of representations of an algebra with four generators
$S_0, S_{\alpha},\ \alpha=1,2,3$,
subject to the homogeneous quadratic relations
\begin{eqnarray}
&&[S_0,S_{\alpha}]_- =
iJ_{\beta \gamma}[S_{\beta}, S_{\gamma}]_+
\nonumber\\
&&[S_{\alpha},S_{\beta}]_-
=i[S_0,S_{\gamma}]_+
\label{skl1}
\end{eqnarray}
Here and below $\{\alpha ,\beta , \gamma \}$ is any {\it cyclic}
permutation of $\{1 ,2,3\}$, $[A,B]_{\pm}=AB\pm BA$.
The structure constants have the form
\beq
J_{\alpha \beta}={J_{\beta}-J_{\alpha}\over J_{\gamma}}
\label{07}
\eeq
where $J_{\alpha}$ are arbitrary parameters.
The algebra generated by the $S_a$ with relations
(\ref{skl1}) and structure constants (\ref{07}) is called
{\it Sklyanin algebra}. There is a
two\--pa\-ra\-met\-ric family of such
algebras. The relations of the Sklyanin algebra imposed
on $S_a$ are equivalent to the condition that the $L$-operator
of the form
\beq
L(u)=\sum_{a=0}^3 W_{a}(u)\,S_{a}\otimes \sigma_{a}
\label{08}
\eeq
(considered as an operator in
${\cal H}\otimes {\bf C}^2$, where ${\cal H}$
is a module over the
algebra) satisfies eq.\,(\ref{02}).
Hence, any finite-dimensional
representation of the Sklyanin algebra
provides a solution to eq.\,(\ref{02}).

As it was shown in \cite{Skl2}, the operators
$S_a,\ a=0,\ldots,3$ admit a realization as
second order difference operators in the space
of meromorphic functions $F(z)$ of a complex variable $z$.
One of the series of such representations
(called the principal analytic series
or series a) in \cite{Skl2}) is
\beq
(S_a F)(z)=
\frac{ (i)^{\delta _{a,2}} \theta _{a+1}(\eta )}
{\theta _{1}(2z)}\Bigl ( \theta _{a+1}(2z\!-\!2\ell \eta )
F(z\!+\!\eta)-\theta _{a+1}(-\!2z\!-\!2\ell \eta )
F(z\!-\!\eta )\Bigr )
\label{repr}
\eeq
(hereafter $\theta (z)\equiv \theta (z|\tau)$).
A straightforward but tedious computation shows that
the operators (\ref{repr}) for any  $\tau,\ \eta, \ \ell$
satisfy the commutation relations
(\ref{skl1}) with the
following values of the structure constants:
\beq
J_{\alpha}={\theta_{\alpha +1}(0)
\theta_{\alpha +1}(2\eta )\over
\theta_{\alpha +1}^2(\eta )}
\label{010}
\eeq
Therefore, $\tau$ and $\eta$ parametrize the structure
constants, while $\ell$ characterizes a representation.

In \cite{kz}, a connection
of the representation theory of the Sklyanin
algebra with the finite-gap theory of soliton equations
was found.
It has been proved that for integer $\ell$ the operator $S_0$ is
algebraically integrable and, therefore,
is a difference analogue of the classical Lame
operator
$$
{\cal L}=-{d^2\over dx^2}+\ell (\ell +1)\wp(x)
$$
which can be obtained
from $S_0$ in the limit $\eta\to 0$.
Finite gap properties of
higher Lame operators (for arbitrary integer values of
$\ell$) were established in \cite{ince}.
Algebraic integrability of $S_0$ implies, in particular,
extremely unusual spectral properties of this operator.
Putting $F_n=F(n\eta+z_0)$,
we assign to (\ref{repr}) the difference Schr\"odinger operators
\beq
S_a F_n=A_n^a F_{n+1}+B_n^a F_{n-1}
\label{14a}
\eeq
with quasiperiodic coefficients. The spectrum of a generic
operator of this form in the space $l^2({\bf Z})$
(square integrable sequences
$F_n$) has a Cantor set type structure. If $\eta$ is a rational
number, $\eta=P/Q$, the operators (\ref{14a})
have $Q$-periodic coefficients.
In general, $Q$-periodic difference Schr\"odinger
operators have $Q$ unstable bands in the spectrum.

It was shown \cite{kz} that
the operator $S_0$ given by eq.\,(\ref{repr})
for positive integer values
of $\ell$ and {\it arbitrary} $\eta$
has $2\ell$ unstable bands in the spectrum.
Its Bloch functions $\psi(z)$ are parametrized by points
of a hyperelliptic curve of genus $2\ell$ defined by the equation
\beq
y^2=P(\varepsilon)=\prod_{i=1}^{2\ell+1}
(\varepsilon^2-\varepsilon_i^2)
\label{15a}
\eeq
Considered as a function
on the curve, $\psi$ is the Baker-Akhiezer function.
Moreover, Bloch eighenfunctions $\psi(z, \pm \varepsilon_i)$
of the operator $S_0$ at
the edges of bands span an invariant functional subspace for all
the operators $S_a$.
The corresponding $4\ell +2$-dimensional representation of
the Sklyanin algebra is a direct sum of two equivalent
$2\ell+1$-dimensional
irreducible representations of the Sklyanin algebra.

As it is well-known from the early days of
the finite-gap theory, the ring of operators commuting with a
finite-gap operator is isomorphic to a ring of meromorphic functions
on the corresponding spectral curve with poles at "infinite
points". For difference operators
it was proved in \cite{mam},\,\cite{kr1}. Therefore, the ring of the
operators commuting with $S_0$
is generated by $S_0$ and an operator $D$ such
that
$$
D^2=P(S_0)=\prod_{i=1}^{2\ell +1}(S_{0}^{2}-\varepsilon_i^2)
$$
In \cite{krnov}, for any algebraic curve
$\G$ of genus $g$ with two punctures,
a special basis in the ring ${\cal M}$
of meromorphic functions $A_i$ with
poles at the punctures was introduced.
These functions define {\it the almost
graded structure} in ${\cal M}$,
i.e. the product of two basis functions has
the form
$$
A_iA_j=\sum_{k=-g/2}^{g/2} c_{i,j}^k A_{i+j+k}
$$
Therefore, for any finite-gap
difference operator $S$ there exist commuting
with $S$ operators $M_i$ such that
$M_i\psi=A_i \psi$,
where $\psi$ is the Baker-Akhiezer eigenfunction of $S$.
The ring generated by the
operators $M_i$ has the same
almost graded structure as the ring ${\cal M}$.
A particular case of the last result corresponding to the ring
of operators commuting with the Sklyanin operator
$S_0$ was recently rediscovered in \cite{varfeld}.

The established connection of the Sklyanin operator
$S_0$ with the algebro-geometric theory of soliton equations is
a part of the theory of integrable multi-dimensional differential or
difference linear operators with elliptic coefficients. It turns
out that the spectral theory of such operators is isomorphic to the
theory of finite-dimensional integrable systems. Among them are
spin generalizations of the Calogero-Moser system,
the Ruijesenaars-Schneider systems and nested Bethe ansatz
equations (see \cite{bab},\,\cite{kz} and \cite{KLWZ},
respectively).

In \cite{kz}, we have suggested a relatively
simple way to derive the realization (\ref{repr}) of
the Sklyanin algebra by difference operators.
Our approach clarifies their origin
as diference operators acting on
the {\it vacuum curve} of the $L$-operator (\ref{08}).
This notion, introduced by one of the authors
\cite{Krichvac}, proved to be useful in
analysis of the Yang-Baxter equation
by methods of algebraic geometry.
The construction of vacuum curve and vacuum vectors
is a suitable generalization of a key property of the elementary
$R$-matrix (\ref{03}), which was used by
Baxter in his famous solution of the 8-vertex model
and called by him "pair-propagation through a
vertex" \cite{Baxter}. In that particular case the vacuum curve
is an elliptic curve.

In this paper we present a more detailed analysis of
the vacuum curve of the higher spin $L$-operator (\ref{08}).
The main result of this work is
a new realization of the
Sklyanin algebra by difference operators acting in {\it two}
variables rather than one (Sects.\,5, 6).
Remarkably, the "finite-gap"
operator $S_0$ in this new realization preserves its form,
while the form of the
other three generators changes drastically, with both variables
entering them in a non-trivial way.
This realization naturally follows from
the explicit construction
of the vacuum curve of the elliptic $L$-operator and
corresponding vacuum vectors.

Let us recall the main definitions.
Consider an {\it arbitrary} $L$-operator $L$
with two-dimensional auxiliary space ${\bf C}^2$, i.e., an arbitrary
$2N\times 2N$ matrix represented as $2\times 2$ matrix whose matrix
elements are $N\times N$ matrices
$L_{11}$, $L_{12}$, $L_{21}$,
$L_{22}$:
\beq
L=\left ( \begin{array}{cc} L_{11}& L_{12}\\
L_{21}& L_{22} \end{array} \right ).
\label{5.0}
\eeq
They act in
a linear space ${\cal H} \cong {\bf C}^N$
which is called the {\it quantum space} of
the $L$-operator.

Let $X\in {\cal H},\, \bigl |U\bigl >=
\left (\begin{array}{c}U_1\\U_2 \end{array}
\right ) \in {\bf C}^2 $ be two vectors
such that
\beq
L\Bigl (X\otimes \bigl |U\bigr > \Bigr )=
Y\otimes \bigl |V\bigl >
\label{5.1a}
\eeq
where $Y\in {\cal H}$, $\bigl |V\bigl >\in {\bf C}^2 $
are some vectors.
Suppose (\ref{5.1a}) holds; then the vector $X$ is called
a {\it vacuum vector} of the $L$-operator. Multiplying
(\ref{5.1a}) from the left by the
covector ${}^{\bot}\!\bigl < V\bigr |=(V_2, -V_1)$,
orthogonal to $\bigl |V\bigr >$,
we get the necessary and sufficient
condition for existence of the vacuum vectors:
\beq
{}^{\bot}\!\bigl < V \bigr |L \bigl |U\bigr >X=0
\label{5.2}
\eeq
Here
${}^{\bot}\!\bigl < V \bigr |L \bigl |U\bigr >$
is an operator in ${\cal H}$.
The {\it vacuum curve} is defined by the equation
\beq
\label{curve}
\mbox{det}\Bigl ({}^{\bot}\!\bigl < V \bigr |L \bigl |U\bigr >
\Bigr )=0
\eeq
By construction, it is embedded into
${\bf C}P^1\times {\bf C}P^1$.

The relation (\ref{5.1a}) (in the particular case ${\cal H}\cong
{\bf C}^2$) was the starting point for Baxter in his solution of
the 8-vertex and $XYZ$ models \cite{Baxter}.
In the context of the
quantum inverse scattering method \cite{FT}
the equivalent condition (\ref{5.2}) is more customary.
It defines local vacua of the
(gauge-transformed) $L$-operator.
A generalization of that
solution to the higher spin $XYZ$ model was given
by Takebe in \cite{Tak1}, where, in particular,
the generalized vacuum vectors were
constructed. However, the vacuum curve itself was implicit
in that work.

In Sect.\,2 we recollect the main formulas related to
the Sklyanin algebra and its representations.
As we shall see in Sect.\,3, the vacuum curves corresponding
to finite-dimensional representations
of the principal analytic series \cite{Skl2} of the
Sklyanin algebra are completely
reducible, i.e. their equations have the form
$$
P(U,V)=\prod_{k=1}^M P_{k}(U,V)=0
$$
with some polynomials $P_k$.
If the dimension of ${\cal H}$ is even,
then $2M=\dim {\cal H}$ and
all irreducible components
are elliptic curves isomorphic to the vacuum
curve of the $R$-matrix $R$ (\ref{03}) found by Baxter.
If the dimension of ${\cal H}$ is odd, then
$2M=\dim {\cal H}+1$ and all but one
irreducible components are isomorphic
to the elliptic curve, while the last component is
rational. We would like to mention, in passing,
that one of the variables in
the new realization of the Sklyanin algebra arises just
as the index $k$ marking irreducible components of the vacuum curve.
The other one is, like in \cite{kz}, a uniformization parameter
on any one of the elliptic components.

The origin of this realization can be traced back to the structure
of the vacuum curve of the operator $\Lambda$ defined by the left
(or right) hand side of eq.\,(\ref{02}).
By definition, $\Lambda$ is an operator in the tensor product
${\cal H}\otimes {\bf C}^2\otimes {\bf C}^2$.
Let us consider it as an
operator in
$\hat {\cal H}\otimes {\bf C}^2$,
where $\hat {\cal H}={\cal H}\otimes
{\bf C}^2$ is product of the first two factors.
Proceeding in the way explained above, we may
assign a vacuum curve to this operator.
As it was shown in \cite{Krichvac}, this
curve is a result of a "composition"
of vacuum curves of the operators $L$ and
$R$. This curve is completely reducible, too. Besides,
all but one components of this latter curve are {\it two-fold
degenerate}, i.e.,
the equation of the curve has the form
$$
\hat P(U,V)= \hat P_0(U,V)
\left [\prod_{i=1}^N \hat P_i(U,V) \right]^2=0
$$
The multiplicity of components of
the composed vacuum curve "mixes" different
components of the vacuum curve of the
$L$-operator and -- in terms of action to the
vacuum vectors -- leads to shifts in the
index $k$ which appear in the two-variable
realization of the Sklyanin generators. A detailed
study of action of the $L(u)$ to the vacuum vectors
is given in Sect.\,4.

At last, in Sect.\,7 we make some remarks on the
trigonometric degeneration of the construction
presented in Sects.\,4-6. This is related to representations
of the quantum algebra $U_q(sl(2))$. In particular,
our approach provides a new type of representations of
this algebra which is a $q$-analogue of representations
of the $sl(2)$ algebra by vector fields on the two-dimensional
sphere.

\section{The elliptic $L$-operator and the Sklyanin algebra}

In this section we collect main formulas related to
the Sklyanin algebra and its representations.

Consider the elliptic $L$-operator (\ref{08}):
\begin{eqnarray}
L(u)&=&
\sum_{a=0}^3 W_{a}(u )
S_{a}\otimes \sigma_{a}
\nonumber\\
&=&\left ( \begin{array}{cc}
W_0 (u)S_0 +W_3 (u)S_3 &
W_1 (u)S_1-iW_2 (u)S_2 \\ \\
W_1 (u)S_1 +iW_2 (u)S_2 &
W_0 (u)S_0-W_3 (u)S_3
\end{array} \right )
\label{L}
\end{eqnarray}
with $W_a(u)$ given by (\ref{2}).
In the sequel we write $\theta _a(x)\equiv \theta _a(x|\tau)$,
$\theta _a(x|\frac{\tau}{2})\equiv \bar \theta _a(x)$
for brevity.
The operators $S_a$ obey Sklyanin algebra (\ref{skl1})
with the structure constants
\beq
J_{\alpha \beta}=-(-1)^{\alpha -\beta}\frac{
\theta_{1}^{2}(\eta)\theta_{\gamma +1}^{2}(\eta)}
{\theta_{\alpha +1}^{2}(\eta)\theta_{\beta +1}^{2}(\eta)}
\label{skl3}
\eeq
We remind that $\{\alpha ,\beta , \gamma \}$ is any cyclic
permutation of $\{1 ,2,3\}$.
Note that the coefficients $W_a$ in (\ref{L})
satisfy the algebraic relations
\beq
(W_{\alpha}^{2}-W_{\beta}^{2})=
J_{\alpha \beta}(W_{\gamma}^{2}-W_{0}^{2})
\label{5}
\eeq
Any two of them are independent and
define an elliptic curve
${\cal E}_{0}\subset {\bf C}P^3 $ as
intersection of the two quadrics. The spectral parameter $u$
uniformizes this curve, i.e., (\ref{5}) is identically satisfied
under the substitutions (\ref{2}), (\ref{skl3}).
Another useful form of the structure
constants is
\beq
J_{\alpha \beta}= \frac{J_{\beta}-J_{\alpha}}{J_{\gamma}}\,,
\;\;\;\;\;\;\;\;
J_{\alpha}={\theta_{\alpha +1}(2\eta)\theta_{\alpha +1}(0)\over
\theta_{\alpha +1}^2(\eta )}
\label{skl4}
\eeq
(see (\ref{07})).
The algebra has two independent central elements:
\beq
\Omega _{0}=\sum _{a=0}^{3}S_{a}^{2}\,,
\;\;\;\;\;\;\;\;
\Omega _{1}=\sum _{\alpha=1}^{3}J_{\alpha}S_{\alpha}^{2}
\label{center}
\eeq

In some formulas below it is
more convenient to deal with the renormalized generators:
\beq
\label{skl5}
S_{a}=(i)^{\delta _{a,2}}\theta _{a+1}(\eta ){\cal S}_a
\eeq
The relations (\ref{skl1}) can be rewritten
in the form
\beq
\begin{array}{l}
(-1)^{\alpha +1}I_{\alpha 0}{\cal S}_{\alpha}{\cal S}_{0}=
I_{\beta \gamma}{\cal S}_{\beta}{\cal S}_{\gamma}
-I_{\gamma \beta}{\cal S}_{\gamma}{\cal S}_{\beta}
\\ \\
(-1)^{\alpha +1}I_{\alpha 0}{\cal S}_0 {\cal S}_{\alpha}=
I_{\gamma \beta}{\cal S}_{\beta}{\cal S}_{\gamma}
-I_{\beta \gamma}{\cal S}_{\gamma}{\cal S}_{\beta}
\end{array}
\label{skl6}
\eeq
where
$$
I_{ab}=\theta _{a+1}(0)\theta _{b+1}(2\eta)
$$
The central elements in the renormalized generators read
\beq
\Omega _{0}=\theta _{1}^{2}(\eta){\cal S}_{0}^{2}+
\sum _{\alpha =1}^{3}(-1)^{\alpha +1}
\theta _{\alpha +1}^{2}(\eta){\cal S}_{\alpha}^{2}\,,
\;\;\;\;\;\;\;\;
\Omega _{1}=\sum _{\alpha=1}^{3}(-1)^{\alpha +1}
I_{\alpha \alpha}{\cal S}_{\alpha}^{2}
\label{center1}
\eeq

Sklyanin's realization \cite{Skl2} of the algebra by difference
operators has the form (\ref{repr}).
The parameter
$\ell$ is "spin" of the representation. When $\ell \in \frac{1}{2}
{\bf Z}_{+}$, these operators have a finite-dimensional invariant
subspace ${\cal T}_{4\ell}^{+}$ of {\it even} $\theta$-functions of
order $4\ell$, i.e., the space
of entire functions $F(z)$, $z\in {\bf C}$,
such that $F(-z)=F(z)$ and
\beq
\begin{array}{l}
F(z+1)=F(z)
\\ \\
F(z+\tau)=\exp (-4\ell \pi i \tau -8\ell \pi i z)F(z)
\end{array}
\label{8}
\eeq
This is the representation space of a $(2\ell +1)$-dimensional
irreducible representation (of series a))
of the Sklyanin algebra\footnote{There are three other series of
irreducible representations \cite{Skl2},\,\cite{smith},\,\cite{kz}
which we do not discuss here.}.
In the spin-$\ell$ represenation of series a),
the central elements take the values
\beq
\begin{array}{l}
\Omega_{0}=4\theta _{1}^{2}
\bigl ( (2\ell +1)\eta \bigr )
\\ \\
\Omega_{1}=4\theta _{1}\bigl (2\ell\eta \bigr )
\theta _{1}\bigl (2(\ell +1)\eta \bigr )
\end{array}
\eeq
In terms of shift operators $T_{\pm}
\equiv \exp (\pm \eta \p _{z})$ (\ref{repr})
can be rewritten for the
renormalized generators (\ref{skl5}) in a little bit simpler
form:
\beq
\label{repr1}
{\cal S}_a =
\frac{\theta _{a+1}(2z-2\ell \eta)}{\theta _{1}(2z)}
\,T_{+}
-\frac{\theta _{a+1}(-2z-2\ell \eta)}{\theta _{1}(2z)}
\,T_{-}
\eeq

Plugging the difference operators
(\ref{repr}) into (\ref{L}), one can represent the $L$-operator
in a "factorized" form \cite{factorized} which is
especially convenient in the computations.
Introduce the matrix
\beq
\hat \Phi (u;z)=\left ( \begin{array}{lll} \bar \theta_4 (z-\ell
\eta +\frac{u}{2}) && \bar \theta_3 (z-\ell \eta +\frac{u}{2})
\\ && \\ \bar
\theta_4 (z+\ell \eta -\frac{u}{2}) && \bar \theta_3 (z+\ell \eta
-\frac{u}{2})\end{array} \right )
\eeq
Then it holds
\beq
L(u)F(z)=2\theta_1(u+2\ell \eta)\hat \Phi ^{-1}(u+4\ell \eta;z)
\left ( \begin{array}{cc}F(z+\eta)&0\\&\\0&F(z-\eta)\end{array}
\right )\hat \Phi (u;z)
\label{fact}
\eeq
where elements of the $L$-operator are assumed to act
to the function $F(z)$ according to (\ref{repr}).

\section{The vacuum curve}

Our goal in this section is to find an explicit
parametrization of the vacuum curve
and vacuum vectors of the elliptic
$L$-operator (\ref{L}) in a finite-dimensional
representation of the Sklyanin algebra.

In practical computations,
it will be more convenient to write 2-dimensional vectors
like $\bigl |U\bigr >$ from the left of the $L$-operator
rather than from the right, so the basic equation (\ref{5.1a})
acquires the form
\beq
\bigl <U\bigr | L(u)\,X =\bigl < V\bigr |\, Y
\label{9}
\eeq
Here $\bigl <U\bigr | = (U_1 , U_2 )$,
$\bigl <V\bigr | = (V_1 , V_2 )$
are {\it covectors}. It is easy to see that
this leads to the definition of the vacuum curve,
$\mbox{det}\bigl (\bigl <U\bigr |
L\bigl |V^{\bot}\bigr >\bigr )=0$,
equivelent to the one given in the Introduction.

Let the operators $S_a$ in (\ref{L}) be
realized as in (\ref{repr}) and let them act in the
finite-dimensional subspace
${\cal T}_{4\ell}^{+}$ (\ref{8}). Then
$X$ and $Y$ in (\ref{9}) are functions of $z$ belonging to the space
${\cal T}_{4\ell}^{+}$.
In this section it is convenient to normalize
$\bigl <U\bigr |$ and
$\bigl <V\bigr |$ by the condition that the second components are
equal to 1: $\bigl <U\bigr | = (U,\,1)$,
$\bigl <V\bigr | =(V,\,1)$.

In this notation the equation of the vacuum curve
(\ref{curve}) acquires the form
\beq
\det K(U,V) =0
\label{10}
\eeq
where
\beq
K(U,V)=(U-V)W_0 S_0 +(1-UV)W_1 S_1 +i(1+UV)W_2 S_2 +(U+V)W_3 S_3
\label{11}
\eeq
In other words, we have to find a relation on $U$ and $V$ under that
the operator $K(U,V)$ has an
eigenvector (belonging to ${\cal T}_{4\ell}^{+}$)
with zero eigenvalue (a "zero mode").
One can write $K(U,V)$ in an equivalent form
$$
K(U,V)= \sum _{a=0}^{3} \zeta _{a}W_a S_a
$$
where the new variables $\zeta _{a}$ are subject to
the constraint
\beq
\zeta _{1}^{2}+\zeta _{2}^{2}+\zeta _{3}^{2}=\zeta _{0}^{2}
\label{constr}
\eeq
\begin{th}
The vacuum curve of the $L$-operator (\ref{L})
for $\ell \in {\bf Z}_{+}
+\frac{1}{2}$
splits into a union of $\ell +\frac{1}{2}$ elliptic curves
isomorphic to ${\cal E}_{0}$ (\ref{5}),
for $\ell \in {\bf Z}_{+}$
the vacuum curve splits into
$\ell$ components isomorphic to ${\cal E}_{0}$
and one rational component.
The equation defining the vacuum curve in the
coordinates $U,\, V$ has the form
\beq
P_{\ell}(U,V)=0,\;\;\;\;\;\ell \in {\bf Z}_{+}+\frac{1}{2}
\label{eq1}
\eeq
\beq
(U-V)^{-1}P_{\ell}(U,V)=0,\;\;\;\;\;\ell \in {\bf Z}_{+}
\label{eq2}
\eeq
where
$$
P_{\ell}(U,V)=\prod _{n=0}^{[\ell ]}
(U^2 +V^2 -2\Delta _{\ell -n}UV +
\Gamma _{\ell -n}(1+U^2 V^2))
$$
($[\ell]$ denotes the integer part of $\ell$),
$\Gamma _{0}=0$, $\Delta _{0}=1$ and the $u$-independent
constants $\Gamma _{k},\, \Delta _{k}$
for $k\geq 1$ are defined below
in (\ref{26}), (\ref{27}).
\end{th}
\paragraph{Remark}
Note that at integer $\ell$ the
polynomial $P_{\ell}(U,V)$ is divisible by $(U-V)^2$, so
the left hand side of (\ref{eq2}) is a polynomial.

\noindent
{\it Proof.} Let us consider first a linear
combination of the generators
$S_a$ with arbitrary coefficients:
\beq
K=\sum _{a=0}^{3}y_a S_a
\label{12}
\eeq
According to (\ref{repr}), the equation $KX=0$ is equivalent to
\beq
\frac{s(-z-\ell \eta )}{ s(z-\ell \eta )}=
\frac{ X(z+\eta)}{X(z-\eta)}
\label{13}
\eeq
where
\beq
s(z)=\sum _{a=0}^{3}(i)^{\delta _{a,2}}y_a
\theta _{a+1} (\eta) \theta _{a+1}(2z)
\label{14}
\eeq
belongs to the space ${\cal T}_4$ of $\theta$-functions of order 4
with the following monodromy properties: $s(z+1)=s(z)$,
$s(z+\tau)=\exp (-4\pi i \tau -8\pi i z )s(z)$. Note that
$\dim {\cal T}_4 =4$ and $s(z)$ has 4 zeros in the fundamental domain
of the lattice formed by 1 and $\tau$. The function $s(z)$ can be
parametrized by its zeros:
\beq
s(z)=\prod _{i=1}^{4}\theta _{1}(z-a_i ),\;\;\;\;\;\;\;
\sum _{i=1}^{4}a_i =0
\label{15}
\eeq
(up to an inessential factor). Functions $\theta _{a+1}(2z)$ form
a basis in the space ${\cal T}_{4}$. Expanding (\ref{15}) with respect
to this basis, we get:
\beq
s(z)=2\sum _{a=0}^{3}(-1)^{a}\theta _{a+1}(2z)
\prod _{i=2}^{4}\theta _{a+1}(-a_1 -a_i )
\label{16}
\eeq

Let us first consider the case $\ell \ge 1$.
Since $X(z)$ has $4\ell$
zeros in the fundamental domain, the equality (\ref{13}) is possible
only if the functions $X(z+\eta)$ and $X(z-\eta)$ have
$4\ell -4$ common zeros. Such a cancellation of zeros of the
numerator and the denominator in the r.h.s. of (\ref{13}) takes
place if zeros of $X(z)$ are arranged into "strings", i.e.,
\beq
X(z)=\prod _{i=1}^{4}
\prod _{j=0}^{m_i -1}\theta _{1}(z-z_i -2j\eta)
\label{17}
\eeq
with the condition
\beq
\sum _{i=1}^{4}m_i =4\ell\,, \;\;\;\;m_i \ge 0
\label{18}
\eeq
At this stage we do not impose any other restrictions; in particular,
it is not implied that $X(z)$ is even.
The number $m_i$ is
called length of the string.
So, if all $m_i >0$, there are four strings
in (\ref{17}) with total length $4\ell$. If some of $m_i$ equal zero,
the number of strings is less than four.
Let $X(z)$ be given by (\ref{17}),
then
\beq
\frac{ X(z+\eta) }{ X(z-\eta) }=
\prod _{i=1}^{4}\frac{ \theta _{1}(z-z_i +\eta) }
{\theta _{1}(z-z_i -(2m_i -1)\eta ) }
\label{19}
\eeq
Identifying zeros of the left hand side of (\ref{13}) with zeros of
(\ref{19}), we get
\beq
z_i +a_i =-(\ell -1)\eta \,, \;\;\;\;i=1, \ldots ,4.
\label{20}
\eeq
Taking these relations into account, we then identify poles of (\ref{13})
and (\ref{19}) and conclude that (unordered) sets of points
$(z_i +(2m_i -1)\eta)$ and $(-z_i +\eta)$,
$i=1, \ldots ,4$ must coincide. In order to describe all the possibilities,
denote by $P$ any permutation of indices (1234) and consider systems
of linear equations
\beq
z_i +z_{P(i)}=-2(m_i -1)\eta\,, \;\;\;\; i=1, \ldots ,4
\label{21}
\eeq
for each $P$. Solutions to these systems consistent with the condition
\beq
\sum _{i=1}^{4} z_i =-4(\ell -1)\eta
\label{22}
\eeq
which follows from (\ref{15}) and (\ref{20}), yield all possible values
of $z_i$. Depending on the choice of $P$ rank of the linear system
(\ref{21}) may equal 4, 3 or 2, i.e., the number of free parameters
in the solutions may be respectively 0, 1 or 2. Since
coefficients of the function $s(z)$ with respect to the basis
$\theta _{a}(2z)$ are already constrained by one relation
(\ref{constr}) (describing the embedding ${\bf C}P^1 \times
{\bf C}P^1 \subset {\bf C}P^3$), the vacuum curve corresponds to the case
of minimal rank. It is easy to see that system (\ref{21}) has rank 2
for the following three permutations of (1234): (2143), (3412), (4321);
otherwise its rank is greater than 2. Solving (\ref{21}) for these
three permutations, we arrive at the following result.

The function $s(z)$ has the form
\beq
s(z)\equiv s^{(m)}(z)=\theta _{1}(z-\mu _{1})\theta _{1}(z-\mu _{2})
\theta _{1}(z\!+\!\mu _{1}\!+\!2(\ell \!-\!m)\eta)
\theta _{1}(z\!+\!\mu _{2}\!-\!2(\ell \!-\!m)\eta )
\label{22a}
\eeq
where $m=0,\, 1,\ldots ,[\ell]$ and
$\mu _{1},\, \mu _{2}$ are free parameters. The corresponding "zero mode"
of $K$ is given by the formula \footnote{These functions coincide with
"intertwining vectors" from the paper \cite{Tak1} found there by
a different method.}
\begin{eqnarray}
X^{(m)}(z) & = &
\prod _{j=1}^{m}
\theta _{1}\left (z+\mu _{1}+(\ell +1-2j)\eta
\right )
\theta _{1}\left (z-\mu _{1}-(\ell +1-2j)\eta
\right ) \nonumber\\
&&\prod _{j=1}^{2\ell -m}\theta _{1}\left
(z+\mu _{2}+(\ell +1-2j)\eta \right )
\theta _{1}\left (z-\mu _{2}-(\ell +1-2j)\eta \right )
\label{zm}
\end{eqnarray}
Note that $X^{(m)}(z)$ is even function.

Let us set $k=\ell \!-\!m$, $k=0,\,1, \ldots, \ell$
for $\ell \in {\bf Z}_{+}$ and $k= \frac{1}{2},\,
\frac{1}{2}, \ldots , \ell$ for
$\ell \in {\bf Z}_{+}\!+\!\frac{1}{2}$ and identify
$\mu _{1}+\mu _{2}=u$ (the spectral parameter),
$\mu _{2}-\mu _{1}=2\zeta +2k\eta$,
where $\zeta$ is a uniformizing parameter
on the vacuum curve. It follows from (\ref{2}), (\ref{14}),
(\ref{16}) and (\ref{22}) that the coefficients $\zeta _a$ in
(\ref{11}) are given by the formula
\beq
\zeta _a =\frac{
2(i)^{\delta _{a,2}}\theta _{a+1}(2k\eta )
\theta _{a+1} (2\zeta )}
{\bar \theta _{3}(\zeta +n\eta )
\bar \theta _{3} (\zeta -k\eta )}
\label{23}
\eeq
They satisfy homogeneous quadratic equations

\beq
\begin{array}{l}
\zeta _{1}^{2}+\zeta _{2}^{2}+\zeta _{3}^{2}=\zeta _{0}^{2}\,,
\\ \\
\displaystyle{ \sum _{\alpha =1}^{3} \zeta _{\alpha}^{2}
\frac{\theta _{\alpha +1}^{2}(0)}
{\theta _{\alpha +1}^{2}(2k\eta )} =0}
\end{array}
\label{24}
\eeq

\noindent
which provide purely algebraic description of the vacuum curve.
Rescaling the coordinates,
$\zeta _{a}=\xi _{a}\theta _{a+1}(2k\eta)$
(for $k\ne 0$), one may represent (\ref{24}) in the form independent
of $k$:

\beq
\begin{array}{l}
\displaystyle{ \sum _{\alpha =1}^{3} \xi _{\alpha}^{2}
\theta _{\alpha +1}^{2}(0)}=0\,, \\ \\
\displaystyle{ \sum _{\alpha =1}^{3} \xi _{3-\alpha}^{2}
\theta _{\alpha +1}^{2}(0)}=0
\end{array}
\label{24a}
\eeq

\noindent
It is straightforward to see that the equations (\ref{5}) defining the
elliptic curve ${\cal E}_{0}$ can be transformed to the same form.
This means that the vacuum curve is reducible: it splits into a union
of components isomorphic to ${\cal E}_0$ (corresponding to nonzero
values of $k$).

In the coordinates $U,V$ the
system (\ref{24}) is equivalent to a single equation of degree 4:
\beq
U^2 +V^2 -2\Delta _{k}UV +\Gamma _{k}(1+U^2 V^2 )=0
\label{25}
\eeq
where
\beq
\Gamma _{k}=
\frac{ \theta _{1}^{2}(2k\eta )\theta _{4}^{2}(2k\eta ) }
{ \theta _{2}^{2}(2k\eta )\theta _{3}^{2}(2k\eta ) }=
\left ( \frac{ \bar \theta _{1}(2k\eta )}
{ \bar \theta _{2}(2k\eta )} \right )^{2}
\label{26}
\eeq
\begin{eqnarray}
\Delta _{k} & = &
\frac{ \theta _{3}^{2}(0) }
{ \theta _{4}^{2}(0) }
\left (
\frac{ \theta _{4}^{2}(2k\eta ) }
{ \theta _{3}^{2}(2k\eta ) }+
\frac{ \theta _{1}^{2}(2k\eta ) }
{ \theta _{2}^{2}(2k\eta ) } \right )
\nonumber\\
&=&
\frac{ \bar \theta _{2}^{2}(0)
\bar \theta _{3}(2k\eta )
\bar \theta _{4}(2k\eta ) }
{ \bar \theta _{2}^{2}(2k\eta)
\bar \theta _{3}(0 )
\bar \theta _{4}(0)}
\label{27}
\end{eqnarray}
For $k=\frac{1}{2}$
this equation after trivial redefinitions coincides with
the equation of the vacuum curve for spin
$\frac{1}{2}$ derived for the first time
by Baxter \cite{Baxter}. So formulas (\ref{24})-(\ref{27}) are valid
for $\ell =\frac{1}{2}$
as well (in this case $k$ takes only one value $\frac{1}{2}$).

The case $k=0$ (4 strings of length $\ell$ each) needs a separate
consideration. In this case $s(z)$ is
an even function, so $\zeta _{0}=0$ and the corresponding component of
the vacuum curve is a rational curve (the cone). Therefore,
for integer $\ell$ the vacuum curve has a rational component
(corresponding to $k=0$) given by the equation $U=V$.
\square

\paragraph{Remark}
At $\tau = 0$ the elliptic
$L$-operator (\ref{L})
degenerates into a trigonometric one. It corresponds to
the higher spin $XXZ$ model.
Its vacuum curve was studied in \cite{kor}.
This curve can be
obtained from our formulas (\ref{25})-(\ref{27})
in the limit $\tau \to 0$, $\eta \to 0$ provided
$\eta '\equiv \eta /\tau$ is finite:
$\Gamma _{k}=0$, $\Delta _k = \cos (4\pi k\eta')$.
Eq.\,(\ref{25}) turns into
$(U-e^{4\pi ik\eta '}V)(U-e^{-4\pi ik\eta '}V)=0$, so
each elliptic component (of degree 4) splits into two rational
component (of degree 2 each).
The equation for the whole vacuum curve
acquires the form
$$
\prod _{n=-[\ell]}^{[\ell]}\left (U-e^{4\pi in\eta '}V\right )=0
$$
Other types of
rational degenerations are also possible \cite{GZ}.
Their detailed classification is not discussed here.

We call the "copy" of the elliptic curve ${\cal E}_0$
defined by eq.\,(\ref{25}) with $k=\ell$ {\it the highest
component} of the vacuum curve.
The results of \cite{kz} suggest that
it plays a distinguished role in representations
of the Sklyanin algebra.

Let us summarize the results of this
section and prepare some formulas
which will be extensively used in the sequel.
The vacuum curve of the elliptic spin-$\ell$ $L$-operator
consists of $[\ell \!+\!\frac{1}{2}]$
components which are marked by
$k=0,1,\ldots ,\ell$ for integer $\ell$ and
$k= \frac{1}{2},  \frac{3}{2}, \ldots ,
\ell$ for half-integer $\ell$. Each elliptic component
is uniformized by the variable $\zeta$:
$$
U=\frac{\bar \theta _{4}(\zeta +k\eta)}
{\bar \theta _{3}(\zeta +k\eta)}\,,
\;\;\;\;\;\;\;\;
V=\frac{\bar \theta _{4}(\zeta -k\eta)}
{\bar \theta _{3}(\zeta -k\eta)}
$$
(at $k=0$ it degenerates into a rational component).
In the sequel it will be more convenient to pass to
the homogeneous coordinates and work with the following
two-dimensional covectors:
\beq
\big <\zeta \big |\equiv
\left ( \bar \theta_4(\zeta ),\,
\bar \theta_3(\zeta ) \right )\,,
\;\;\;\;\;\;\;\;
\big <-\zeta \big |=\big <\zeta \big |
\eeq
The vacuum vectors are given by
(\ref{zm}), which we rewrite in terms of $\zeta$, $k$:

\begin{eqnarray}
\label{X}
X_{k}^{\ell}(z,\zeta )&=&
\prod _{j=1}^{\ell -k}
\theta_1\!\left (z\!-\!\zeta \!+\!\frac{u}{2}\!+\!
(\ell \!-\!k \!+\!1 \!-\!2j)\eta \right )
\theta_1 \!\left (z\!+\!\zeta \!-\!\frac{u}{2}-
(\ell \!-\!k \!+\!1 \!-\!2j)\eta \right )
\nonumber \\
&&\prod _{j=1}^{\ell +k}\theta_1
\!\left (z\!+\!\zeta \!+\!\frac{u}{2}\!+\!
(\ell \!+\!k \!+\!1\! -\!2j)\eta \right )
\theta_1 \!\left (z\!-\!\zeta \!-\!\frac{u}{2}\!-\!
(\ell \!+\!k \!+\!1 \!-\!2j)\eta \right )
\end{eqnarray}

\noindent
We call $X_{\ell}^{\ell }$ {\it the highest vacuum vector}.
The formula (\ref{X}) still has sense for negative
values of $k$, too, extending
$X_{k}^{\ell}(z,\zeta)$ to $-\ell \leq k \leq \ell$.
From now on we assume that $k$ varies in this region.
Let us point out the following simple properties
of vacuum vectors:

$$
\begin{array}{l}
X_{k}^{\ell}(-z, \zeta)=
X_{k}^{\ell}(z, \zeta)
\\ \\
X_{k}^{\ell}(z, -\zeta)=
X_{-k}^{\ell}(z, \zeta)
\\ \\
X_{k}^{\ell}(z+\eta, \zeta)=
X_{k}^{\ell}(z, \zeta +\eta)
\displaystyle{\frac{\theta_1(z-\zeta +\frac{u}{2}-
(k -\ell)\eta )\theta_1(z-\zeta -\frac{u}{2}+(k+\ell)\eta )}
{\theta_1(z-\zeta +\frac{u}{2}+
(k -\ell)\eta )\theta_1(z-\zeta -\frac{u}{2}-(k+\ell)\eta )}}
\\ \\
X_{k}^{\ell}(z, \zeta +\eta)=
X_{k+1}^{\ell}(z, \zeta )
\displaystyle{\frac{\theta_1(z-\zeta +\frac{u}{2}-
(\ell -k)\eta )\theta_1(z+\zeta -\frac{u}{2}+(\ell -k)\eta )}
{\theta_1(z+\zeta +\frac{u}{2}-
(\ell +k)\eta )\theta_1(z-\zeta -\frac{u}{2}+(\ell +k)\eta )}}
\end{array}
$$
In the next section we study how $L(u)$ acts to the vacuum
vectors.

\section{Action of $L(u)$ to the vacuum vectors}

By a straightforward computation we obtain:
\beq
\label{LXY}
\big <\zeta +k\eta \big |\,L(u)X_{k}^{\ell}(z,\zeta)=
2\theta_1 (u-2\ell \eta)\big <\zeta -k\eta \big |\,
Y_{k}^{\ell}(z,\zeta +\eta)
\eeq
where
\begin{eqnarray}
Y_{k}^{\ell}(z,\zeta)&=&
\prod _{j=0}^{\ell -k-1}
\theta_1 \left (z\!-\!\zeta \!+\!\frac{u}{2}\!+\!
(\ell \!-\!k \!+\!1 \!-\!2j)\eta \right )
\theta_1\!\left (z\!+\!\zeta \!-\!\frac{u}{2}\!-\!
(\ell \!-\!k \!+\!1 \!-\!2j)\eta \right )
\nonumber \\
&&\!\!\prod _{j=1}^{\ell +k}
\theta_1 \left (z\!+\!\zeta \!+\!\frac{u}{2}\!+\!
(\ell \! +\!k \!+\!1 \!-\!2j)\eta \right )
\theta_1 \!\left (z\!-\!\zeta \!-\!\frac{u}{2}\!-\!
(\ell \!+\!k \!+\!1\! -\!2j)\eta \right )
\end{eqnarray}
It is easy to see that
$$
Y_{k}^{\ell}(z,-\zeta)=
Y_{-k}^{\ell}(z,\zeta \!+\!2\eta)
$$
Indicating explicitly the vacuum vector dependence on
the spectral parameter $u$, we get for any
$-\ell \leq k \leq\ell$:
\beq
\label{YX}
Y_{k}^{\ell}(z,\zeta \!+\!\eta ,u)=
X_{k}^{\ell}(z,\zeta ,u \!+\!2\eta)
\eeq
and
$$
X_{-k}^{\ell}(z,\zeta ,u)=
X_{k}^{\ell}(z,\zeta ,-u)
$$
Below we supress the $u$-dependence of vacuum vectors if it
is not misleading.

For the highest vacuum vectors we have
$$
Y_{\ell}^{\ell }(z,\zeta)= X_{\ell}^{\ell}(z,\zeta)
$$
Therefore, at $k=\ell$ one can rewrite
eq.\,(\ref{LXY}) in the form

\beq
\label{LXX}
\big <\zeta \pm \ell \eta \big |\,L(u)
X_{\pm \ell}^{\ell}(z,\zeta)=
2\theta_1 (u-2\ell \eta)\big <\zeta \mp \ell \eta \big |\,
X_{\pm \ell}^{\ell}(z,\zeta \pm \eta)
\eeq

\noindent
The equation with the lower sign is obtained by
changing $\zeta \to -\zeta$ and using the above listed
properties of the vacuum vectors.
From (\ref{X}) we see that
\beq
\label{XuX}
X_{\ell}^{\ell}\Bigl (z,\, \zeta \!-\!\frac{u}{2},\,u \Bigr )=
X_{-\ell}^{\ell}\Bigl (z,\, \zeta \!+\!\frac{u}{2},\,u \Bigr )
\eeq
This property allows us to convert (\ref{LXX}) into a closed
system of equations for the vector $X_{\ell}^{\ell}$ only.
Indeed, let us substitute $\zeta \to \zeta \mp \frac{u}{2}$ in
the first (second) equality in (\ref{LXX}) and after that
make use of (\ref{XuX}). In this way we get the following
system of equations:
\beq
\label{LXPM}
\begin{array}{l}
\big <\zeta -\frac{u}{2}+ \ell \eta \big |\,L(u)
X_{\ell}^{\ell}(z,\zeta -\frac{u}{2})=
2\theta_1 (u-2\ell \eta)\big <\zeta -\frac{u}{2}-\ell \eta \big |\,
X_{\ell}^{\ell}(z,\zeta -\frac{u}{2}+\eta)
\\ \\
\big <\zeta +\frac{u}{2}- \ell \eta \big |\,L(u)
X_{\ell}^{\ell}(z,\zeta -\frac{u}{2})=
2\theta_1 (u-2\ell \eta)\big <\zeta +\frac{u}{2}+\ell \eta \big |\,
X_{\ell}^{\ell}(z,\zeta -\frac{u}{2}-\eta)
\end{array}
\eeq
In our paper \cite{kz}
it is shown that the representation
(\ref{repr}) in the variable $\zeta$
follows from solution to this system.
Therefore, we can say that these representations are realized
in the space of functions on the highest component of the
vacuum curve.

Now let us turn to other components of the vacuum curve.
It is possible to express action of the $L$-operator
in terms of the vectors
$X_{k}^{\ell}(z,\zeta )$ only. A straightforward
computation leads to the following result:

\begin{eqnarray}
&&\Bigl <\zeta \!\pm \!w \!+\! k\eta \Bigr |\,L(u\!-\!2w)
X_{k}^{\ell}(z,\zeta ,u)
\nonumber\\
&&
\nonumber\\
&=&2\,\displaystyle{\frac{
\theta_1 (2(\zeta\mp \ell \eta ))
\theta_1 (u-2w\mp 2k\eta )}
{\theta_1 (2(\zeta-k \eta ))}}
\,\Bigl <\zeta \!\pm \!w \!-\!k\eta
\Bigr |X_{k}^{\ell}(z,\zeta \pm \eta,u)
\nonumber\\
&&
\nonumber\\
&\pm &2\,\displaystyle{\frac{
\theta_1 (2(k\mp \ell) \eta )
\theta_1 (2\zeta \mp u\pm 2w)}
{\theta_1 (2(\zeta-k \eta ))}}
\,\Bigl <\zeta \!\mp \!w \!-\!k\eta
\Bigr |X_{k\pm 1}^{\ell}(z,\zeta,u)
\label{LXXX}
\end{eqnarray}

\noindent
Here $w$ is an arbitrary parameter.
The formulas with upper and lower signs are connected
by the transformation $\zeta \to -\zeta$ and $k \to -k$.
Equalities of this type are sometimes called "intertwining
relations" or "vertex-face correspondence". In the particular
case $\ell =\frac{1}{2}$ they were suggested by Baxter
\cite{Baxter} and recently generalized to any half-integer
values of $\ell$ by Takebe \cite{Tak1}.

Putting $w$ equal to $0$ and comparing
these formulas with (\ref{LXY}), we get the following linear
relations between vectors $Y_{k}^{\ell}$ and $X_{k}^{\ell}$:

\begin{eqnarray}
Y_{k}^{\ell}(z,\zeta +\eta,u)&=&
\displaystyle{\frac{
\theta_1 (2(\zeta \mp \ell \eta ))
\theta_1 (u \mp 2k\eta )}
{\theta_1 (2(\zeta-k \eta ))
\theta_1 (u-2\ell \eta ))}}
\,X_{k}^{\ell}(z,\zeta \pm \eta,u)
\nonumber\\
&&
\nonumber\\
&\pm &\displaystyle{\frac{
\theta_1 (2(k\mp \ell) \eta )
\theta_1 (2\zeta \mp u)}
{\theta_1 (2(\zeta-k \eta ))
\theta_1 (u-2\ell \eta ))}}
\,X_{k\pm 1}^{\ell}(z,\zeta,u)
\label{YXX}
\end{eqnarray}

\noindent
The two choices of signs in the r.h.s.
provide a 4-term linear relation between vectors
$X_{k}^{\ell}$ among themselves.
Moreover, taking into account the property (\ref{YX}),
one can rewrite (\ref{YXX}) in a more symmetric form

\beq
\begin{array}{ll}
&\theta_1 (2(\zeta-k \eta ))
\theta_1 (u-2\ell \eta )
\,X_{k}^{\ell}(z,\zeta,u\!+\!2\eta)
\\ &\\
=&\theta_1 (2(\zeta-\ell \eta ))
\theta_1 (u-2k \eta )
\,X_{k}^{\ell}(z,\zeta \!+\!\eta ,u)
\\
+&\theta_1 (2(k-\ell)\eta )
\theta_1 (2\zeta -u)
\,X_{k+1}^{\ell}(z,\zeta,u)
\\ &\\
=&\theta_1 (2(\zeta+\ell \eta ))
\theta_1 (u+2k \eta )
\,X_{k}^{\ell}(z,\zeta \!-\!\eta ,u)
\\
-&\theta_1 (2(k+\ell)\eta )
\theta_1 (2\zeta +u)
\,X_{k-1}^{\ell}(z,\zeta,u)
\end{array}
\label{XXX}
\eeq

\noindent
Note that the three variables $\zeta$, $k$ and $u$
after a proper rescaling enter symmetrically inspite of their
very different nature.

\section{Difference operators on the vacuum curve}

Formulas
(\ref{LXXX}) give rise to some distinguished
difference operators in
{\it two} variables -- $\zeta$ and $k$ --
related to representations of the Sklyanin algebra.
Consider a particular case of the first
equation (\ref{LXXX}) (with upper sign) at $u=0$, $w=-\frac{u}{2}$.
From (\ref{X}) it follows that $X_{k}^{\ell}(z,\zeta,0)$ is
{\it even function of $\zeta$}. Note that it automatically
implies
$X_{-k}^{\ell}(z,\zeta,0)=X_{k}^{\ell}(z,\zeta,0)$.
Therefore, the
substitution $\zeta \to -\zeta$ provides us with another
equation for the same vector $X_{k}^{\ell}(z,\zeta,0)$.
Together with the first one they form a closed system:

\begin{eqnarray}
&&\Bigl <\zeta \!\mp \!\frac{u}{2} \pm k\eta \Bigr |\,L(u)
X_{k}^{\ell}(z,\zeta ,0)
\nonumber\\
&&
\nonumber\\
&=&2\,\displaystyle{\frac{
\theta_1 (2(\zeta\mp \ell \eta ))
\theta_1 (u-2k\eta )}
{\theta_1 (2(\zeta \mp k \eta ))}}
\,\Bigl <\zeta \!\mp \frac{u}{2} \mp \!k\eta
\Bigr |X_{k}^{\ell}(z,\zeta \pm \eta,0)
\nonumber\\
&&
\nonumber\\
&+&2\,\displaystyle{\frac{
\theta_1 (2(k-\ell) \eta )
\theta_1 (2\zeta \mp u)}
{\theta_1 (2(\zeta \mp k \eta ))}}
\,\Bigl <\zeta \!\pm \!\frac{u}{2}\! \mp \!k\eta
\Bigr |X_{k+1}^{\ell}(z,\zeta,0)
\label{LXXX0}
\end{eqnarray}

\noindent
which is an extension of the simpler system
(\ref{LXPM}) on the highest component to the whole
vacuum curve.

Let $X_i$, ${}^{i}\!\bar X$,
$i=1,2,\ldots , 2\ell +1$, be
dual bases in the space
${\cal T}_{4\ell}^{+}$ of theta-functions and its dual
${\cal T}_{4\ell}^{+,*}$, respectively, i.e.
$({}^{i}\!\bar X, \, X_j)=\delta _{ij}$, where
$(\,\, ,\,\,)$ denotes pairing of the spaces
${\cal T}_{4\ell}^{+,*}$ and ${\cal T}_{4\ell}^{+}$.
The linear space ${\cal F}_{\zeta, k}$ of functions
spanned by
$${}^{i}\!X^{\ell}_{k}(\zeta)\equiv
\Bigl ( {}^{i}\!\bar X,\, X^{\ell}_{k}(z,\zeta,0)\Bigr )
$$
which are functions of $\zeta$, $k$ but not of $z$, plays
a central role in what follows.
Let ${\cal A}$ be any difference operator in $z$. One can
translate its action to the space
${\cal F}_{\zeta, k}$ according to the definition
\beq
\label{def}
\Bigl ({\cal A}\circ {}^{i}\!X^{\ell}_{k}\Bigr )(\zeta)=
\Bigl ({}^{i}\!\bar X,\,{\cal A}X^{\ell}_{k}(z,\zeta,0)\Bigr )
\eeq
This action is extendable to the whole space
${\cal F}_{\zeta, k}$ by linearity.

Note the composition rule
\beq
{\cal A}\circ ({\cal A}' \circ F)=
({\cal A}'{\cal A})\circ F
\label{comp}
\eeq
for any $F\in {\cal F}_{\zeta ,k}$, i.e. the order of the
operators ${\cal A}$, ${\cal A}'$ with respect to the
action $\circ$ must be reversed. Indeed, let us write
$$
X_{k}^{\ell}(z,\zeta )=\sum _{i}{}^{i}\!X^{\ell}_{k}(\zeta)
X_i(z)
$$
and define matrix elements of an operator ${\cal A}$ to be
$$
({\cal A})^{j}_{i}=
\Bigl ( {}^{i}\!\bar X,\, {\cal A}
X_{i}(z)\Bigr )
$$
Then we have
$$
{\cal A}\circ {}^{j}\!X^{\ell}_{k}=\sum _{i}
({\cal A})^{j}_{i}\,\,{}^{i}\!X^{\ell}_{k}\,,
\;\;\;\;\;\;\;
{\cal A}'\circ {}^{j}\!X^{\ell}_{k}=\sum _{i}
({\cal A}')^{j}_{i}\,\,{}^{i}\!X^{\ell}_{k}\,,
$$
$$
\begin{array}{lll}
({\cal A}'{\cal A})\circ {}^{j}\!X^{\ell}_{k}&=&
\Bigl ({}^{j}\!\bar X,\,
{\cal A}'{\cal A}X^{\ell}_{k}\Bigr )
\\ && \\
&=& \displaystyle{\sum _{i,l}} ({\cal A}')^{j}_{l}
({\cal A})^{l}_{i}\,\,{}^{i}\!X^{\ell}_{k}
\\ && \\
&=&{\cal A}\circ ({\cal A}' \circ {}^{j}\!X^{\ell}_{k})
\end{array}
$$

After these preliminaries we can
make a convolution of the system (\ref{LXXX0}) with
the basis covectors ${}^{i}\bar X$ and rewrite it
in terms of the action $\circ$ as a relation between
functions in the space ${\cal F}_{\zeta, k}$:

\begin{eqnarray}
&&\Bigl <\zeta \!\mp \!\frac{u}{2} \pm k\eta \Bigr |\,L(u)
\circ X_{k}^{\ell}(\zeta )
\nonumber\\
&&
\nonumber\\
&=&2\,\displaystyle{\frac{
\theta_1 (2(\zeta\mp \ell \eta ))
\theta_1 (u-2k\eta )}
{\theta_1 (2(\zeta \mp k \eta ))}}
\,\Bigl <\zeta \!\mp \frac{u}{2} \mp \!k\eta
\Bigr |X_{k}^{\ell}(\zeta \pm \eta)
\nonumber\\
&&
\nonumber\\
&+&2\,\displaystyle{\frac{
\theta_1 (2(k-\ell) \eta )
\theta_1 (2\zeta \mp u)}
{\theta_1 (2(\zeta \mp k \eta ))}}
\,\Bigl <\zeta \!\pm \!\frac{u}{2}\! \mp \!k\eta
\Bigr |X_{k+1}^{\ell}(\zeta)
\label{LXXX1}
\end{eqnarray}

\noindent
(here and below $X_{k}^{\ell}(\zeta )
\in {\cal F}_{\zeta ,k}$).
These relations form a system of four linear equations
for the four functions $(S_a \circ X^{\ell}_{k})(\zeta)$
entering the left hand side. To make this clear,
it is useful to
rewrite the system in a more explicit form:

$$
\left ( \begin{array}{lll} \bar \theta_4 (\zeta\!-\!
\frac{u}{2}\!+\!k\eta ) && \bar \theta_3 (z\!-\!
\frac{u}{2}\!+\!k\eta )
\\ && \\ \bar
\theta_4 (z\!+\!\frac{u}{2}\!-\!k\eta ) &&
\bar \theta_3 (z\!+\!
\frac{u}{2}\!-\!k\eta )\end{array} \right )
\left ( \begin{array}{lll}
L_{11}(u)\circ X_{k}^{\ell}(\zeta ) &&
L_{12}(u)\circ X_{k}^{\ell}(\zeta )
\\ && \\
L_{21}(u)\circ X_{k}^{\ell}(\zeta ) &&
L_{22}(u)\circ X_{k}^{\ell}(\zeta )
\end{array} \right )
$$

$$
= \,2\,
\left ( \begin{array}{lll} W_{11}\bar \theta_4 (\zeta\!-\!
\frac{u}{2}\!-\!k\eta )
X_{k}^{\ell}(\zeta +\eta )
&& W_{11}\bar \theta_3 (z\!-\!
\frac{u}{2}\!-\!k\eta )X_{k}^{\ell}(\zeta +\eta )
\\ && \\ W_{22}\bar
\theta_4 (z\!+\!\frac{u}{2}\!+\!k\eta )
X_{k}^{\ell}(\zeta -\eta )
&& W_{22}\bar \theta_3 (z\!+\!
\frac{u}{2}\!+\!k\eta )X_{k}^{\ell}(\zeta -\eta )
\end{array} \right )
$$

$$
+ \,2\,
\left ( \begin{array}{lll} W_{12}\bar \theta_4 (\zeta\!+\!
\frac{u}{2}\!-\!k\eta ) && W_{12}\bar \theta_3 (z\!+\!
\frac{u}{2}\!-\!k\eta )
\\ && \\ W_{21}\bar
\theta_4 (z\!-\!\frac{u}{2}\!+\!k\eta ) &&
W_{21}\bar \theta_3 (z\!-\!
\frac{u}{2}\!+\!k\eta )\end{array} \right )
X_{k+1}^{\ell}(\zeta)
$$

\noindent
where $W_{ij}$ are elements of the matrix
$$
W= \left ( \begin{array}{lll}
\displaystyle{\frac{ \theta _{1}(2(\zeta -\ell \eta ))
\theta _{1}(u-2k\eta )}
{\theta _{1}(2(\zeta -k\eta ))}}
&&
\displaystyle{\frac{ \theta _{1}(2(k -\ell )\eta )
\theta _{1}(2\zeta -u )}
{\theta _{1}(2(\zeta -k\eta ))}}
\\ && \\
\displaystyle{\frac{ \theta _{1}(2(k -\ell )\eta )
\theta _{1}(2\zeta +u )}
{\theta _{1}(2(\zeta +k\eta ))}}
&&
\displaystyle{\frac{ \theta _{1}(2(\zeta +\ell \eta ))
\theta _{1}(u-2k\eta )}
{\theta _{1}(2(\zeta +k\eta ))}}
\end{array} \right )
$$

\noindent
Solving this system, we
obtain the following action rules of the generators $S_a$
to functions of two variables:

\beq
\label{S0}
(S_0 \circ X_{k}^{\ell})(\zeta)\!=\!
\frac{\theta_1(\eta)}{\theta_1(2\zeta)}\!\Bigl (
\theta_1(2\zeta\!-\!2\ell \eta)X_{k}^{\ell}(\zeta \!+\!\eta)\!+\!
\theta_1(2\zeta\!+\!2\ell
\eta)X_{k}^{\ell}(\zeta \!-\!\eta) \!\Bigr )
\eeq

\begin{eqnarray}
\label{Si}
(S_{\alpha}\circ X_{k}^{\ell})(\zeta)\!&=&
\!-(i)^{\delta _{a,2}}
\frac{\theta_{\alpha +1}(\eta)}
{\theta_{1}(2\zeta)}\left (
\frac{\theta_1(2\zeta -2\ell \eta)
\theta_{\alpha +1}(2\zeta -2k\eta)}
{\theta_{1}(2\zeta -2k\eta)}
X_{k}^{\ell}(\zeta +\eta) \right.
\nonumber\\
&&
\nonumber\\
&-&\left. \frac{\theta_1(2\zeta +2\ell \eta)
\theta_{\alpha +1}(2\zeta +2k\eta)}
{\theta_{1}(2\zeta +2k\eta)}
X_{k}^{\ell}(\zeta -\eta) \right )
\nonumber\\
&&
\nonumber\\
&+&
\!\!(i)^{\delta _{a,2}}
\frac{2\theta_{\alpha +1}(\eta)
\theta_{\alpha +1}(2k\eta)
\theta_{\beta +1}(2\zeta)
\theta_{\gamma +1}(2\zeta)
\theta_{1}\bigl (2 (k \!-\!\ell )\eta \bigr)}
{\theta_{\beta +1}(0)
\theta_{\gamma +1}(0)
\theta_{1}\bigl (2\zeta \!-\!2k\eta \bigr )
\theta_{1}\bigl (2\zeta \!+\!2k\eta \bigr )}
X_{k\!+\!1}^{\ell}(\zeta)
\end{eqnarray}

\noindent
Note that the formula for $S_0$ remains the same as in
Sklyanin's realization while the other generators
are substantially different. They are non-trivial operators
in two variables rather than one.

\paragraph{Remark}
At $k=\ell$ the last term in the r.h.s.
of (\ref{Si}) disappears and
we come back to Sklyanin's formulas in the variable $\zeta$.
However, the last three generators
differ from the ones in (\ref{repr}) by a sign.
This sign can be explained if we recall that the action
$\circ$ has "contravariant" composition rule (\ref{comp})
that amounts to the formal transpositon of the commutation
relations. Commutation relations of the Sklyanin algebra
imply that the transposition of all generators is equivalent
(up to an automorphism of the algebra) to
changing signs of $S_1$, $S_2$ and $S_3$. Hence formulas
(\ref{S0}), (\ref{Si}) at $k=\ell$
are indeed equivalent to (\ref{repr}).

Do the difference operators
(\ref{S0}), (\ref{Si}) obey the
(transposed) Sklyanin algebra?
The answer is {\it no} since, as it is clear from (\ref{Si}),
the highest shifts in $k$ do not cancel in the commutation relations
(\ref{skl6}).
The matter is that functions $X_{k}^{\ell}(\zeta)$ are
not arbitrary functions of two variables.
By construction, they belong to
the space ${\cal F}_{\zeta , k}$. This implies an additional
condition that follows from (\ref{YXX}) at $u=0$
after convolution with the basis covectors ${}^{i}\bar X$:

\begin{eqnarray}
\label{cond}
&&\frac{\theta_1(2\zeta-2\ell\eta)}
{\theta_1(2\zeta )}X_{k}^{\ell}(\zeta +\eta)
+\frac{\theta_1(2\zeta+2\ell\eta)}
{\theta_1(2\zeta )}X_{k}^{\ell}(\zeta -\eta)
\nonumber\\
&&
\nonumber\\
&=&\frac{\theta_1(2k\eta-2\ell\eta)}
{\theta_1(2k\eta )}X_{k+1}^{\ell}(\zeta )
+\frac{\theta_1(2k\eta+2\ell\eta)}
{\theta_1(2k\eta )}X_{k-1}^{\ell}(\zeta )
\end{eqnarray}

\noindent
This equality means that $S_0$ has a "dual" realization as a
difference operator in the variable $k$ which has the same form.

Investigating commutation properties of the
difference operators standing in
(\ref{S0}), (\ref{Si}), it is convenient to modify them
in two respects. First, let us change signs of
$S_{\alpha}$, thus coming back to the "covariant" action
(see the remark above). Second, it is natural to
disregard the origin of the
$k$, $\ell$ as (half) integer numbers and allow them to
take arbitrary complex values.
Namely, let us set
$x=k\eta$, $K_{\pm}=\exp (\pm \eta \p _{x})$,
$T_{\pm}=\exp (\pm \eta \p _{\zeta})$.
In this notation
our difference operators
in two variables $\zeta$, $x$ are
\beq
{\cal D}_0 = \frac{\theta _{1}(2\zeta -2\ell \eta)}
{\theta _{1}(2\zeta)}\, T_{+}
+\frac{\theta _{1}(2\zeta +2\ell \eta)}
{\theta _{1}(2\zeta)}\, T_{-}
\label{S0a}
\eeq
\begin{eqnarray}
\label{Sia}
{\cal D}_{\alpha}&=&
\frac{\theta _{1}(2\zeta -
2\ell \eta)\theta _{\alpha +1}(2\zeta -2x)}
{\theta _{1}(2\zeta) \theta _{1}(2\zeta -2x)}\,
T_{+}\!-\!
\frac{\theta _{1}(2\zeta +2\ell \eta)
\theta _{\alpha +1}(2\zeta +2x)}
{\theta _{1}(2\zeta) \theta _{1}(2\zeta +2x)}\,
T_{-}
\nonumber\\
&-&
\frac{2\theta_{\alpha +1}(2x)
\theta_{\beta +1}(2\zeta)
\theta_{\gamma +1}(2\zeta)
\theta_{1}\bigl (2x -2\ell \eta \bigr)}
{\theta_{\beta +1}(0)
\theta_{\gamma +1}(0)
\theta_{1}\bigl (2\zeta \!-\!2x \bigr )
\theta_{1}\bigl (2\zeta \!+\!2x \bigr )}\,
K_{+}
\end{eqnarray}
The last line can be transformed as follows:
\begin{eqnarray}
\label{Siaa}
{\cal D}_{\alpha}&=&
\frac{\theta _{1}(2\zeta -2\ell \eta )}
{\theta _{1}(2\zeta )}
\frac{\theta _{\alpha +1}(2\zeta -2x)}
{\theta _{1}(2\zeta -2x)}\,T_{+}
-
\frac{\theta _{1}(2\zeta +2\ell \eta )}
{\theta _{1}(2\zeta )}
\frac{\theta _{\alpha +1}(2\zeta +2x)}
{\theta _{1}(2\zeta +2x)}\,T_{-}
\nonumber\\
&-&
\frac{\theta _{1}(2x -2\ell \eta )}
{\theta _{1}(2x )}
\left (
\frac{\theta _{\alpha +1}(2\zeta -2x)}
{\theta _{1}(2\zeta -2x)}
-\frac{\theta _{\alpha +1}(2\zeta +2x)}
{\theta _{1}(2\zeta +2x)}\right )K_{+}
\end{eqnarray}
In the next section we show that these operators do form a
non-standard realization
of the Sklyanin algebra on a subspace of functions of two
variables.

\section{Representations of the Sklyanin algebra}

Let us consider the operator
\begin{eqnarray}
\nabla &=&
\frac{\theta _{1}(2\zeta -2\ell \eta)}
{\theta _{1}(2\zeta)}\, T_{+}
+\frac{\theta _{1}(2\zeta +2\ell \eta)}
{\theta _{1}(2\zeta)}\, T_{-}
\nonumber\\
&-&
\frac{\theta _{1}(2x -2\ell \eta)}
{\theta _{1}(2x)}\, K_{+}
-\frac{\theta _{1}(2x +2\ell \eta)}
{\theta _{1}(2x)}\, K_{-}
\label{R1}
\end{eqnarray}
The condition (\ref{cond}) is then
$\nabla X_{k}^{\ell}(\zeta)=0$.
The main statement of this section is:
\begin{th}
For any complex parameter $\ell$
the operators ${\cal D}_a$ (\ref{S0a}), (\ref{Sia})
form a representation of the Sklyanin algebra (\ref{skl6})
in the invariant subspace of functions
of two variables $X=X(\zeta, x)$ such that
$\nabla X =0$. The values of the central elements (\ref{center})
in this representation are
\beq
\begin{array}{l}
\Omega _{0}=4\theta _{1}^{2}\bigl ( (2\ell +1)\eta \bigr )
\\ \\
\Omega _{1}=4\theta _{1}\bigl ( 2\ell \eta \bigr )
\theta _{1}\bigl ( 2(\ell +1)\eta \bigr )
\end{array}
\label{R2}
\eeq
\end{th}

First of all let us show that the space of solutions
to the equation $\nabla X =0$ is invariant under action
of the operators ${\cal D}_a$.
\begin{lem}
The following commutation relations hold:
\beq
\nabla {\cal D}_{a}={\cal D}'_{a}\nabla
\label{R3}
\eeq
where ${\cal D}'_0 ={\cal D}_0$ and
\begin{eqnarray}
\label{R4}
{\cal D}'_{\alpha}&=&
\frac{\theta _{1}(2\zeta -2\ell \eta )}
{\theta _{1}(2\zeta )}
\,T_{+}\,\frac{\theta _{\alpha +1}(2\zeta -2x)}
{\theta _{1}(2\zeta -2x)}
-
\frac{\theta _{1}(2\zeta +2\ell \eta )}
{\theta _{1}(2\zeta )}
\,T_{-}\,\frac{\theta _{\alpha +1}(2\zeta +2x)}
{\theta _{1}(2\zeta +2x)}
\nonumber\\
&-&
\frac{\theta _{1}(2x -2\ell \eta )}
{\theta _{1}(2x )}
\,K_{+}\left (
\frac{\theta _{\alpha +1}(2\zeta -2x)}
{\theta _{1}(2\zeta -2x)}
-\frac{\theta _{\alpha +1}(2\zeta +2x)}
{\theta _{1}(2\zeta +2x)}\right )
\end{eqnarray}
\end{lem}
\begin{cor}
The condition
$$\nabla X=0$$ is invariant under action of
the operators ${\cal D}_a$.
\end{cor}
The lemma can be proved by
straightforward though quite long computations. Let us show
how to reduce them to a reasonable amount.
Our strategy is to begin with the case $\ell =0$.
For brevity we use the notation
\beq
b_{\alpha}(\zeta )=\frac{\theta _{\alpha +1}(2\zeta)}
{\theta _{1}(2\zeta)}
\label{R5}
\eeq
At $\ell =0$ the operators $\nabla$, ${\cal D}_{a}$,
${\cal D}'_{a}$ take the form

\beq
\label{R6}
\nabla =T_{+}+T_{-}-K_{+}-K_{-}
\eeq

\beq
\label{R7}
{\cal D}_{0}={\cal D}'_{0}=T_{+}+T_{-}
\eeq

\beq
\label{R8}
{\cal D}_{\alpha}=b_{\alpha}(\zeta -x)T_{+}
-b_{\alpha}(\zeta +x)T_{-}-\Bigl ( b_{\alpha}(\zeta -x)
-b_{\alpha}(\zeta +x) \Bigr )K_{+}
\eeq

\beq
\label{R9}
{\cal D}'_{\alpha}=T_{+}b_{\alpha}(\zeta -x)
-T_{-}b_{\alpha}(\zeta +x)-K_{+}\Bigl ( b_{\alpha}(\zeta -x)
-b_{\alpha}(\zeta +x) \Bigr )
\eeq

\noindent
It is not too difficult to verify the relations (\ref{R3})
by the direct substitution of these operators.
Note that in this case the specific form of the function
$b_{\alpha}(\zeta )$ given by (\ref{R5}) is irrelevant.
This proves the lemma at $\ell =0$.

The relation $\nabla {\cal D}_{0}={\cal D}'_{0}\nabla$
for general values of $\ell$ is obvious.
To prove the other ones in the case $\ell \neq 0$, we modify the
shift operators as

\beq
\tilde T_{\pm}=c_{\ell}(\pm \zeta )T_{\pm}\,,
\;\;\;\;\;\;\;\;
\tilde K_{\pm}=c_{\ell}(\pm x)K_{\pm}
\label{R10}
\eeq
where
\beq
c_{\ell}(\zeta )=\frac{\theta _{1}(2\zeta -2\ell \eta)}
{\theta _{1}(2\zeta)}
\label{R11}
\eeq

\noindent
These $\tilde T_{\pm}$, $\tilde K_{\pm}$ possess the same
commutation relations as
$T_{\pm}$, $K_{\pm}$ {\it except} for the properties
$T_{+}T_{-}=T_{-}T_{+}=1$ (and similarly for $K_{\pm}$).
The latter are substituted by
\beq
\label{R12}
\tilde T_{\pm}\tilde T_{\mp}=\rho _{\ell}(\pm \zeta )\,,
\;\;\;\;\;\;\;\;
\tilde K_{\pm}\tilde K_{\mp}=\rho _{\ell}(\pm x)
\eeq
where

\beq
\label{R13}
\rho _{\ell}(\zeta)\equiv c_{\ell}(\zeta )
c_{\ell}(-\zeta -\eta )=
\frac{\theta _{1}(2\zeta -2\ell \eta)
\theta _{1}(2\zeta +2(\ell +1)\eta)}
{\theta _{1}(2\zeta )\theta _{1}(2\zeta +2\eta )}
\eeq

\noindent
Clearly, the operators (\ref{R6})-(\ref{R9})
with the substitutions
$T_{\pm} \to \tilde T_{\pm}$,
$K_{\pm} \to \tilde K_{\pm}$ convert into the
corresponding operators for $\ell \neq 0$.
Moreover, it is easy to see that
the same is true for the operator
parts of the both sides of eq.\,(\ref{R3}).
Only the $c$-number contributions get modified.
Collecting them together, we come to the relation

$$
\nabla {\cal D}_{\alpha} ={\cal D}'_{\alpha}\nabla
+f_{\alpha }(\zeta , x)
$$

\noindent
where the function $f_{\alpha}$ is given by
$$
f_{\alpha }\!=\!(\rho _{\ell}(x)\!-\!\rho _{\ell}(\zeta ))
b_{\alpha}(\zeta \!+\!x \!+\!\eta )
\!+\!(\rho _{\ell}(-x)\!-\!\rho _{\ell}(\zeta ))
b_{\alpha}(\zeta \!-\!x \!+\!\eta )
+ (\zeta \,\to  \,-\zeta )
$$
It is easy to verify
that this function has no singularities and has the same
monodromy properties in $\zeta$ as $b_{\alpha}(\zeta )$.
(This time the explicit form of the functions $b_{\alpha}$
is of course crucial.)
Therefore, $f_{\alpha}(\zeta ,x)=0$ and the assertion is proved.

Now let us turn to the commutation relations of the
Sklyanin algebra. It is more convenient to
deal with the relations in the form (\ref{skl6}).
We set
\beq
\label{R14}
\begin{array}{l}
G_{\alpha}\equiv I_{\beta \gamma}
{\cal D}_{\beta} {\cal D}_{\gamma}
-I_{\gamma \beta}{\cal D}_{\gamma} {\cal D}_{\beta}
+(-1)^{\alpha}I_{\alpha 0}{\cal D}_{\alpha}{\cal D}_{0}
\\ \\
\bar G_{\alpha}\equiv I_{\gamma \beta}
{\cal D}_{\beta} {\cal D}_{\gamma}
-I_{\beta \gamma }{\cal D}_{\gamma} {\cal D}_{\beta}
+(-1)^{\alpha}I_{\alpha 0}{\cal D}_{0}{\cal D}_{\alpha}
\end{array}
\eeq

\beq
\label{R15}
\begin{array}{l}
\tilde \Omega_{0}=\theta _{1}^{2}(\eta){\cal D}_{0}^{2}+
\displaystyle{\sum _{\alpha=1}^{3}}(-1)^{\alpha +1}
\theta _{\alpha +1}^{2}(\eta) {\cal D}_{\alpha}^{2}
\\ \\
\tilde \Omega_{1}=
\displaystyle{\sum _{\alpha =1}^{3}}(-1)^{\alpha +1}
I_{\alpha \alpha}{\cal D}_{\alpha}^{2}
\end{array}
\eeq
As is clear from (\ref{skl6}), (\ref{center1}), if
${\cal D}_{\alpha}$ were generators of the Sklyanin algebra,
one would have $G_{\alpha}=\bar G_{\alpha}=0$,
$\tilde \Omega _{0}=\Omega _{0}$,
$\tilde \Omega _{1}=\Omega _{1}$.
\begin{lem}
For any $\ell \in {\bf C}$ it holds

\beq
\label{R16}
\begin{array}{l}
G_{\alpha}=\lambda _{\alpha}(\zeta , x)c_{\ell}(x)K_{+}\nabla
\\ \\
\bar G_{\alpha}=-\lambda _{\alpha}
(\zeta , -x-\eta )c_{\ell}(x)K_{+}\nabla
\end{array}
\eeq

\beq
\label{R17}
\begin{array}{l}
\tilde \Omega_{0}=4\theta _{1}^{2}
\bigl ( (2\ell +1)\eta \bigr )+\lambda '(\zeta , x)
c_{\ell}(x)K_{+}\nabla
\\ \\
\tilde \Omega_{1}=4\theta _{1}\bigl (2\ell\eta \bigr )
\theta _{1}\bigl (2(\ell +1)\eta \bigr )+
\lambda _{0}(\zeta , x)c_{\ell}(x)K_{+}\nabla
\end{array}
\eeq
where
\beq
\label{R18}
\lambda _{a}(\zeta , x) =-2(-1)^{a}b_{a}(\zeta)
b_{a}(x)
\frac{\theta _{1}(2x)\theta _{1}(2x+2\eta)
\theta _{1}(2\zeta)\theta _{1}(2\zeta +2\eta)}
{\theta _{1}(2\zeta -2x)\theta _{1}(2\zeta +2x +2\eta)}
+(\zeta \to -\zeta )
\eeq
\beq
\label{R19}
\lambda '(\zeta , x) =-2\,
\frac{\theta _{1}(2x)\theta _{1}(2x+2\eta)
\theta _{1}^{2}(2\zeta +\eta )}
{\theta _{1}(2\zeta -2x)\theta _{1}(2\zeta +2x +2\eta)}
+(\zeta \to -\zeta )
\eeq

\end{lem}
Again, the computations
can be essentially simplified by dealing with
the case $\ell =0$ first and making the substitutions
(\ref{R10}) after that. Then only $c$-number contributions
need some additional attention.
A few identities used in the computation
are given in the Appendix.

It follows from (\ref{R16}) that in the space of solutions
to the equation $\nabla X=0$ we have
$G_{\alpha}=\bar G_{\alpha}=0$ that proves the theorem.

Let us mention the relation
\beq
{\cal D}'_{\alpha} =-\Xi \,
\check {\cal D}_{\alpha}^{\dagger}\,\Xi ^{-1}
\label{R20}
\eeq
where $\dagger$ means transposition of the operator,

$$
\begin{array}{lll}
\check {\cal D}_{\alpha}&=&b_{\alpha}(\zeta -x)
c_{\ell}(-\zeta -\eta) T_{+}
-b_{\alpha}(\zeta +x)
c_{\ell}(\zeta -\eta) T_{-}\\ &&\\
&-&
\Bigl ( b_{\alpha}(\zeta -x)-
b_{\alpha}(\zeta +x)\Bigr )
c_{\ell}(-x -\eta) K_{+}\end{array}
$$

\noindent
and $\Xi$ is the operator changing the sign of $x$:
$\Xi f(x)=f(-x)$, $\Xi ^{2}=1$.
Note that the operators $\check {\cal D}_{\alpha}$
differ from ${\cal D}_{\alpha}$ by a "gauge" transformation
of the form $U(\zeta , x)(...)U^{-1}(\zeta , x)$ with a function
$U(\zeta , x)$. This transformation acts as follows:
$$
\begin{array}{l}
c_{\ell}(\mp \zeta -\eta)T_{\pm}\,\, \longrightarrow \,\,
c_{\ell}(\pm \zeta )T_{\pm} \\ \\
c_{\ell}(\mp x -\eta)K_{\pm}\,\, \longrightarrow \,\,
c_{\ell}(\pm x )K_{\pm}
\end{array}
$$
and takes $\nabla ^{\dagger}$, ${\cal D}_{0}^{\dagger}$
to $\nabla$, ${\cal D}_{0}$ respectively.
This fact allows us to reduce the computation of $\bar G_{\alpha}$
to that of $G_{\alpha}$.

\paragraph{Remark}
The theorem remains true for operators
$$
\tilde {\cal D}_{a}={\cal D}_{a}+g_a (\zeta , x)\nabla
$$
where $g_a$ are arbitrary functions. Using this remark,
one can write down many other equivalent realizations
of the Sklyanin algebra in two variables. In particular,
it is possible to "symmetrize" the ${\cal D}_a$, i.e.
to include in it all the four shift operators $T_{\pm}$,
$K_{\pm}$ in a symmetric fashion. We do not know
whether it is possible to choose $g_a$ in such a way that
the r.h.s. of (\ref{R16}) would vanish.

\section{Remarks on the trigonometric limit}

The construction of this paper admits several
trigonometric degenerations. Let us outline
the simplest one -- when the
Sklyanin algebra degenerates into the
quantum algebra
$U_{q}(sl(2))$ \cite{KR}-\cite{FRT}.
In this section a {\it multiplicative} parametrization
is more convenient than the {\it additive} one used
in the previous sections. To ease the comparison with
the previous formulas, we denote variables like
$u$, $z$, etc by the same letters having in mind, however,
that shifts are now "multiplicative":
$$
T_{\pm}f(\zeta )=f(q^{\pm1}\zeta )\,,
\;\;\;\;\;\;\;\;
K_{\pm}f(x)=f(q^{\pm1}x)
$$

The $L$-operator reads
\beq
L(u)=\left (
\begin{array}{ccc}
uA-u^{-1}D && (q-q^{-1})C \\ && \\
(q-q^{-1})B && uD -u^{-1}A
\end{array}
\right )
\label{T1}
\eeq
where $A,B,C,D$ are generators of the $U_{q}(sl(2))$:
\beq
\label{T2}
\begin{array}{l}
AB=qBA\,, \;\;\;\;
BD=qDB\,, \;\;\;\;
DC=qCD\,, \;\;\;\;
CA=qAC \\ \\
AD=1\,,
\;\;\;\;\;\;\;\;
BC-CB=\displaystyle{\frac{A^2 -D^2}{q-q^{-1}}}
\end{array}
\eeq
The central element is
\beq
\label{center2}
\Omega =\frac{q^{-1}A^2 +qD^2 }{(q-q^{-1})^2}+BC
\eeq
The standard realization of this algebra by difference operators
has the form
\beq
\begin{array}{l}
(AF)(z)=F(qz)\,, \;\;\;\;\;\;\;\;
(DF)(z)=F(q^{-1}z)
\\ \\
(BF)(z)=\displaystyle{\frac{z}{q-q^{-1}}}\Bigl (q^{\ell}
F(q^{-1}z)-q^{-\ell}F(qz)\Bigr )
\\ \\
(CF)(z)=\displaystyle{\frac{z^{-1}}{q-q^{-1}}}\Bigl (q^{\ell}
F(qz)-q^{-\ell}F(q^{-1}z)\Bigr )
\end{array}
\label{T3}
\eeq
The invariant subspace of the spin-$\ell$ representation is
spanned by $z^{-\ell}, \ldots , z^{\ell}$.
The analogue of the formula (\ref{fact}) has the form
\beq
L(u)F(z)=\bigl (q^{\ell}u-q^{-\ell}u^{-1}\bigr )
\hat \Phi ^{-1}(uq^{2\ell};z)
\left ( \begin{array}{cc}q^{-\ell}F(qz)&0\\&\\0&
q^{\ell}F(q^{-1}z)\end{array}
\right )\hat \Phi (u;z)
\label{fact1}
\eeq
with the matrix
$$
\hat \Phi (u;z)=\left ( \begin{array}{lll}
1 && \bar q^{\ell}u^{-1}z^{-1}
\\ && \\
1 && q^{-\ell}uz^{-1}
\end{array} \right )
$$
The vacuum vectors are (cf. (\ref{X}))
\beq
\label{T4}
X_{k}^{\ell}(z,\zeta)=z^{-\ell}
\prod _{j=1}^{\ell -k}\Bigl (z-\zeta u^{-1}
q^{\ell -k+1-2j}\Bigr )
\prod _{j=1}^{\ell +k}\Bigl (z-\zeta u
q^{\ell +k+1-2j}\Bigr )
\eeq
where $\zeta$ parametrizes the (rational) vacuum curve.

Skipping all the intermediate steps (which are parallel
to the elliptic case),
we turn right to the representation
of the $U_q (sl(2))$ obtained in this way.
One arrives at the following realization of the quantum
algebra by difference operators in two variables $\zeta$, $x$:

\beq
\begin{array}{l}
A=q^{-\ell}T_{+}\,, \;\;\;\;\;\;\;\;
D=q^{\ell}T_{-}
\\ \\
B=\displaystyle{\frac{\zeta}{q-q^{-1}}}\Bigl (q^{\ell}x
T_{-}-q^{-\ell}x^{-1}T_{+}
-\bigl (q^{-\ell}x-q^{\ell}x^{-1}\bigr) K_{+}\Bigr )
\\ \\
C=\displaystyle{\frac{\zeta^{-1}}{q-q^{-1}}}\Bigl (q^{-\ell}
xT_{+}-q^{\ell}x^{-1}T_{-}
-\bigl (q^{-\ell}x-q^{\ell}x^{-1}\bigr) K_{+}\Bigr )
\end{array}
\label{T5}
\eeq

\noindent
This is to be compared with (\ref{S0a}), (\ref{Sia}).
An easy computation shows that
these operators obey the algebra (\ref{T2})
{\it without any additional conditions}! However, in our
present set-up there {\it is}, too, an analogue of
the operator $\nabla$:
\beq
\nabla =
q^{-\ell}T_{+}
+q^{\ell}T_{-}-
\frac{q^{-\ell}x-q^{\ell}x^{-1}}{x-x^{-1}}\,K_{+}
-\frac{q^{\ell}x-q^{-\ell}x^{-1}}{x-x^{-1}}\,K_{-}
\label{T6}
\eeq
and the condition $\nabla X =0$.
Now its role is to restrict the functional space
to the representation space of spin $\ell$.
Indeed, computing the
central element (\ref{center2}), we get:
\beq
\label{T7}
\Omega = \frac{q^{2\ell +1}+q^{-2\ell -1}}{(q-q^{-1})^2}
-\frac{(q^{-\ell}x-q^{\ell}x^{-1})(qx-q^{-1}x^{-1})}
{(q-q^{-1})^2}\,K_{+}\nabla
\eeq
The invariance of the space of solutions to the equation
$\nabla X =0$ is then obvious.

At last, let us discuss continuum
limit ($q \to 1$) of the obtained
formulas. In this limit we arrive at representations
of the algebra $sl(2)$. In fact there are several
different limits $q\to 1$. The most interesting one
reads
\beq
\begin{array}{l}
s_0 =\zeta \p _{\zeta}-\ell \\ \\
s_{+}= \frac{1}{2}\zeta
\Bigl ( (x+x^{-1})(2\ell -\zeta \p _{\zeta})
-(x^2 -1)\p _{x}\Bigr ) \\ \\
s_{-}= \frac{1}{2}\zeta ^{-1}
\Bigl ( (x+x^{-1})\zeta \p _{\zeta}
-(x^2 -1)\p _{x}\Bigr )
\end{array}
\label{T8}
\eeq
where $s_0$, $s_{\pm}$ are standard generators of
the $sl(2)$. After the change of variables
$\zeta =e^{i\varphi }$,
$x=i\,\mbox{cot}\,(\theta /2)$
eqs.\,(\ref{T8}) acquire the form
\beq
\begin{array}{l}
s_0 =-i\p _{\varphi}-\ell \\ \\
s_{+}= e^{i\varphi}(i \p _{\theta}-
\mbox{cot}\,\theta \,\p _{\varphi} +2i\ell \,\mbox{cot}\,\theta )
\\ \\
s_{-}= e^{-i\varphi}(i \p _{\theta}+
\mbox{cot}\,\theta \,\,\p _{\varphi})
\end{array}
\label{T9}
\eeq
At $\ell =0$ this is the well known
realization of $sl_2$ by
vector fields on the two-dimensional sphere.
The representation (\ref{T5}) at $\ell =0$
is its $q$-deformation, with the $q$-deformed variables
being $e^{i\varphi}$ and
$\mbox{cot}\,(\theta /2)$. Another $q$-deformed version
of (\ref{T9}) (at $\ell =0$), where
$e^{i\varphi}$ and $e^{i\theta}$ are "discretized",
has been suggested in \cite{GrZh}.

\section*{Acknowledgenents}
We thank S.Khoroshkin and M.Olshanetsky for useful discussions.
The work of I.K. was supported in part by RFBR grant 95-01-00755.
The work of A.Z. was supported in part by RFBR grant 98-01-00344
and by grant 96-15-96455 for support of scientific schools.

\section*{Appendix}
\addcontentsline{toc}{section}{Appendix}
\def\theequation{A\arabic{equation}}
\setcounter{equation}{0}

We use the following definition of the $\theta$-functions:
\beq
\begin{array}{l}
\theta _1(z|\tau)=\displaystyle{\sum _{k\in {\bf Z}}}
\exp \left (
\pi i \tau (k+\frac{1}{2})^2 +2\pi i
(z+\frac{1}{2})(k+\frac{1}{2})\right ),
\\ \\
\theta _2(z|\tau)=\displaystyle{\sum _{k\in {\bf Z}}}
\exp \left (
\pi i \tau (k+\frac{1}{2})^2 +2\pi i
z(k+\frac{1}{2})\right ),
\\ \\
\theta _3(z|\tau)=\displaystyle{\sum _{k\in {\bf Z}}}
\exp \left (
\pi i \tau k^2 +2\pi i
zk \right ),
\\ \\
\theta _4(z|\tau)=\displaystyle{\sum _{k\in {\bf Z}}}
\exp \left (
\pi i \tau k^2 +2\pi i
(z+\frac{1}{2})k\right )
\end{array}
\label{theta}
\eeq
Throughout the paper we write
$\theta (z|\tau)=\theta (z)$,
$\theta (z|\frac{\tau}{2})=\bar \theta (z)$.

The list of identities used in the computations is given
below.

\beq
\begin{array}{l}
\bar \theta _4 (x)\bar \theta _3(y)+\bar
\theta _4 (y)\bar \theta _3(x)=
2\theta _4 (x+y)\theta_4 (x-y)
\\ \\
\bar \theta _4 (x)\bar \theta _3(y)-
\bar \theta _4 (y)\bar \theta _3(x)=
2\theta _1 (x+y)\theta_1 (x-y)
\\ \\
\bar \theta _3 (x)\bar \theta _3(y)+
\bar \theta _4 (y)\bar \theta _4(x)=
2\theta _3 (x+y)\theta_3 (x-y)
\\ \\
\bar \theta _3 (x)\bar \theta _3(y)-
\bar \theta _4 (y)\bar \theta _4(x)=
2\theta _2 (x+y)\theta_2 (x-y)
\end{array}
\label{theta34}
\eeq

\begin{eqnarray}
&&\bar \theta _{4}(z-x+\ell \eta)
\theta _{1}(z+y+x -(k+\ell )\eta)
\theta _{1}(z-y+x +(k-\ell )\eta)
\nonumber\\
&-&
\bar \theta _{4}(z+x-\ell \eta)
\theta _{1}(z-y-x +(k+\ell )\eta)
\theta _{1}(z+y-x -(k-\ell )\eta)
\nonumber\\
&=&\bar \theta _{4}(y-k \eta)
\theta _{1}(2z)
\theta _{1}(2x -2\ell \eta)
\end{eqnarray}

\noindent
(and the same with the change $\bar \theta _{4}\to
\bar \theta _{3}$)

\begin{eqnarray}
&&\theta _{\alpha +1}(2\zeta -u+2x)\theta _{1}(2\zeta -2x)
\theta _{1}(2\zeta +u)
\nonumber\\
&-&\theta _{\alpha +1}(2\zeta +u-2x)\theta _{1}(2\zeta +2x)
\theta _{1}(2\zeta -u)
\nonumber\\
&=&\frac{\theta _{1}(4\zeta)\theta _{\alpha +1}(2x)}
{\theta _{\alpha +1}(2\zeta )}
\,\theta _{1}(u-2x)\theta _{\alpha +1}(u)\,,
\;\;\;\;\;\;\;\alpha =1,2,3
\end{eqnarray}

\beq
\theta _{1}(2z)
\theta _{2}(0)\theta _{3}(0)\theta _{4}(0)
=2\theta _{1}(z)\theta _{2}(z)\theta _{3}(z)\theta _{4}(z)
\eeq

In the main text we use the notation
$$
\begin{array}{l}
I_{ab}=\theta _{a+1}(0)\theta _{b+1}(2\eta )
\\ \\
b_{\alpha }(\zeta)=\displaystyle{\frac{
\theta _{\alpha +1}(2\zeta)}{\theta _{1}(2\zeta)}}
\\ \\
c_{\ell}(\zeta )=
\displaystyle{\frac{
\theta _{1}(2\zeta -2\ell \eta )}{\theta _{1}(2\zeta)}}
\\ \\
\rho _{\ell}(\zeta )=
c_{\ell}(\zeta ) c_{\ell}(-\zeta -\eta )
\end{array}
$$

\noindent
For verifying the comutation relations (\ref{R3}),
(\ref{R16}) we need the identities

\beq
I_{\gamma \beta }\theta _{\gamma +1}(x)
\theta _{\beta +1}(x\!+\!2\eta )
\!-\!I_{\beta \gamma}\theta _{\beta +1}(x)
\theta _{\gamma +1}(x\!+\!2\eta )\!=\!(-1)^{\alpha}
I_{\alpha 0}\theta _{\alpha +1}(x)
\theta _{1}(x\!+\!2\eta )
\eeq

\beq
b_{\alpha}(\zeta -x)-
b_{\alpha}(\zeta +x)=
2\,\frac{\theta _{\alpha +1}(2x)\theta _{\beta +1}(2\zeta )
\theta _{\gamma +1}(2\zeta )\theta _{1}(2x)}
{\theta _{\beta +1}(0)\theta _{\gamma +1}(0)
\theta _{1}(2\zeta -2x)\theta _{1}(2\zeta +2x)}
\eeq

\beq
\rho _{\ell}(\zeta)
-\rho _{\ell}(x)=
\frac{\theta _{1}(2\ell \eta )
\theta _{1}(2(\ell +1)\eta )
\theta _{1}(2\zeta -2x)\theta _{1}(2\zeta +2x+2\eta )}
{\theta _{1}(2x)\theta _{1}(2x+2\eta )
\theta _{1}(2\zeta )\theta _{1}(2\zeta +2\eta )}
\eeq

\noindent
To find the r.h.s. of eqs.\,(\ref{R17}), we need the
identities

\beq
\sum _{\alpha =1}^{3}(-1)^{\alpha}
\theta _{\alpha +1}^{2}(\eta)
b_{\alpha}(x)b_{\alpha}(x+\eta )=\theta _{1}^{2}(\eta)
\eeq

\beq
\sum _{\alpha =1}^{3}(-1)^{\alpha}
\theta _{\alpha +1}(0)
\theta _{\alpha +1}(2\eta)
b_{\alpha}(x)b_{\alpha}(x+\eta )=0
\eeq

\begin{eqnarray}
&&
\sum _{\alpha =1}^{3}(-1)^{\alpha }
\theta _{\alpha +1}(0)
\theta _{\alpha +1}(x)\theta _{\alpha +1}(y)
\theta _{\alpha +1}(z)
\nonumber\\
&&\nonumber\\
&=&2\theta _{1}\Bigl (\frac{x+y+z}{2}\Bigr )
\theta _{1}\Bigl (\frac{x-y+z}{2}\Bigr )
\theta _{1}\Bigl (\frac{x+y-z}{2}\Bigr )
\theta _{1}\Bigl (\frac{x-y-z}{2}\Bigr )
\end{eqnarray}

\end{document}